\documentclass{aa}

\usepackage{graphicx}
\usepackage{natbib}
\usepackage{pifont}
\usepackage{txfonts}
\usepackage{mathrsfs}
\usepackage{epsfig}
\usepackage{longtable}

\bibpunct{(}{)}{;}{a}{}{,}      

\begin{document}
  \DeclareGraphicsExtensions{.pdf,.mps,.png,.ps,.eps,.jpg}
%
\title{Dust in brown dwarfs and extra-solar planets}
\subtitle{V. Cloud formation in carbon- and oxygen-rich environments}
\author{ }
\author{Ch. Helling        
        \and D. Tootill    
        \and P. Woitke
        \and G. Lee}
 \institute{Centre for Exoplanet Science, SUPA, School of Physics and Astronomy, University of St. Andrews, North Haugh, St. Andrews, Fife, United Kingdom, KY16 9SS\\ email: ch80@st-and.ac.uk}
\date{Accepted 2012 ?,  Received 2012 ?,   in original from \today}

\abstract{Recent observations indicate potentially carbon-rich
  (C/O$>$1) exoplanet atmospheres.  Spectral fitting methods for brown
  dwarfs and exoplanets have invoked the C/O ratio as additional
  parameter but carbon-rich cloud formation modeling is a challenge
  for the models applied. The determination of the habitable zone for
  exoplanets requires the treatment of cloud formation in chemically
  different regimes.
}
{We aim to model cloud formation processes for carbon rich
  exoplanetary atmospheres.  Disk models show that carbon-rich or
  near-carbon-rich niches may emerge and cool carbon planets may trace
  these particular stages of planetary evolution.}
{We extend our kinetic cloud formation model by including carbon seed
  formation and the formation of C[s], TiC[s], SiC[s], KCl[s], and
  MgS[s] by gas-surface reactions.  We solve a system of dust moment
  equations and element conservation for a pre-scribed {{\sc
      Drift-Phoenix }} atmosphere structure to study how a cloud
  structure would change with changing initial C/O$_0$=0.43
  $\,\ldots\,10.0$.}
{The seed formation efficiency is lower in carbon-rich atmospheres
  than in oxygen-rich gases due to carbon being a very effective
  growth species. The consequence is that less particles will make up
  a cloud for C/O$_0\!\!>$1. The cloud particles will be smaller in
  size than in an oxygen-rich atmosphere. An increasing initial C/O
  ratio does not revert this trend because a much greater abundance of
  condensible gas species exists in a rich carbon environment. Cloud
  particles are generally made of a mix of materials: carbon dominates
  if C/O$_0\!\!>$1 and silicates dominate if C/O$_0\!\!<$1. 80-90\%
  carbon is reached only in extreme cases where C/O$_0$=3.0 or 10.0.}
{Carbon-rich atmospheres would form clouds that are made of particles
  of height-dependent mixed compositions, sizes and numbers. The
  remaining gas-phase is far less depleted than in an oxygen-rich
  atmosphere. Typical tracer molecules are HCN and C$_2$H$_2$ in
  combination with a featureless, smooth continuum due to a
  carbonaceous cloud cover, unless the cloud particles become
  crystalline.}


\maketitle


\section{Introduction}


Spectral observations of extrasolar planets reveal their chemical
content which is linked to the planet's formation process but which is
altered by the planetary evolution. Migration and disk evolution,
however, may challenge the direct and easy linking of present-days
observable composition and planet formation. A couple of observations
point to the possibility of planets with carbon-rich
atmospheres. \cite{nikku2011} suggested the WASP-12b has a carbon-rich
atmosphere based on a day-side emission spectrum, despite its host
star being oxygen-rich. This is based on a synthetic spectrum
comparison derived for a parameterised atmosphere structure to fit the
observations.  \cite{krei2015} retrieve from HST  transmission
  spectrum of WASP-12b the presence of H$_2$O  with a
  C/O$\approx0.5$ on the planet's day-side in contrast to
  \cite{nikku2011}, and suggest an atmosphere with clouds and haze
and absorption by e.g. CO, CH$_4$, HCN, C$_2$H$_2$, FeH.  The planets
HD 8988b,c,d,e have tentatively been suggested to show molecules (b \&
d: CH$_4$, C$_2$H$_2$, CO$_2$; c: C$_2$H$_2$; e: CH$_4$, C$_2$H$_2$;
\citealt{op2013}) typical for carbon-rich atmospheres
(\citealt{hell1996}). \cite{tsia2016} analyze WFC3 HST data for the
super-Earth 55 Cancri e and derive a C/O=1.1 on this evaporating,
probably hydrogen-rich planet. A tentative HCN detection is
suggested. \cite{jura2015} report on a polluted Keck/HIRES spectrum of
the white dwarfs WD Ton 345 that suggest that the accreted disk
planetisimals (debris of a disrupted planet) must have been
carbon-rich and water-poor.

Disk chemistry simulations suggest that the local disk gas can become
carbon-rich but only under special circumstances and at certain times
(\citealt{hell2014,ali2014,eis2016}). This suggests that carbon-rich planets
could be a tracer of these rare episodes in an originally oxygen-rich
disk.  When planets form by the core-accretion scenario, smaller
planetary masses would favor an enrichment with heave elements inside
the core. 

The dominating source of carbon (and Al and Li) in the universe are
evolved stars that undergo the third dredge up on the asymptotic giant
branch. Dust-driven winds (radiation pressure due to large 
luminosities) then enrich the ISM with these elements. \cite{abia2003}
show that solar-metallicity AGB stars with C/O$\sim$1, and a C/O$>$2
are considered extreme. Nuclear synthesis and evolutionary models show
that only low-metallicity AGB stars would produce C/O=10
(\citealt{abia2003}). Given the low abundance of carbon-rich,
nearly-sun-like stars (\citealt{fort2012}), carbon-rich planets around
those stars most likely must be linked to the disk evolution rather
than to a primordial C/O$>$1. The situation may differ for planets
around carbon-rich AGB stars which may well have been enriched by the
dust that leaves the star in abundance during its wind phases with
mass losses of $10^{-7}\,\ldots\,10^{-4}$M$_{\odot}$ yr$^{-1}$
(e.g. \citealt{schroe1998,win2003}). The dominating source for carbon
in the early universe, however, are Population III SNe
ejecta. Carbon-rich planets could now form in situ as by-product of
carbon-enhanced metal-poor stars (CEMPs), i.e. carbon-rich Pop II
stars, as suggested by \cite{mas2016}. Would such planets still have
clouds? First investigations for metal-deficient ultra-cool
atmospheres were done for oxygen-rich gases only but demonstrated that
cloud do prevail to unexpected low metallicities of [M/H]$\approx$-5.0
(\citealt{witte2009}).

Linking spectra of extrasolar planets to planet formation and disk
evolution requires therefore the possibility to treat oxygen-rich and
carbon-rich atmospheres and the chemical transition from the one to
the other.  An essential part of this task is to be able to treat
cloud formation in such chemically diverse environments.  Also
determining the habitable zone for extrasolar planets demands detailed
cloud models across various chemical regimes
(\citealt{yan2013,barn2016}).

We present results from a kinetic cloud formation model treating seed
formation, growth/evaporation, gravitational settling and element
depletion to form cloud particles of height-dependent, mixed
compositions and size in carbon-rich and oxygen-rich environments. We
present first results from our next generation of cloud formation
models which follow the moment approach presented in
\cite{woi2003,woi2004,hell2006,hell2008b}.  Section~\ref{sec:method}
summarizes our approach. Section~\ref{CtoOdiff} presents our results
for clouds in atmospheric environments of changing C/O ratios from
C/O$<$1 (oxygen-rich) to C/O$>$1 (carbon-rich), and presents how
global cloud properties change with evolving C/O. One of the questions to
be answered is under which conditions cloud particles would form
primarily ($>80\%$) of pure carbon.  Sections~\ref{s:elm} and
~\ref{sec:molabund} present the results for the remaining gas-phase
elements and discuss some example molecules,
respectively. Section~\ref{s:sum} summarizes this paper.


\section{Approach}\label{sec:method}
We present a new generation of our kinetic, non-equilibrium cloud
formation model which allows us to investigate cloud structures in
oxygen-rich and carbon-rich environments, and the transition between
the two. We assess the impact of the carbon-to-oxygen ratio (C/O) on
the resulting cloud structure details for giant gas planet
atmospheres.  We utilize example model atmosphere structures (T$_{\rm
  eff}$ = 1600 K and 2000K, log(g) = 3.0, initial solar metallicity)
from the {\sf Drift-Phoenix} atmosphere grid that is representative
for the atmosphere of a giant gas planet.  We use the model (T$_{\rm
  gas}$, p$_{\rm gas}$)-structure as input for our external cloud
formation program {\sc Drift} to derive the cloud structures details
for different C/O-ratios  by adjusting the carbon-abdundance. All
  other elements are kept at solar values.  We present our
results for a selected set of C/O ratios (0.43, 1.0, 1.1, 1.5).  We
assume that the planets form via the core-accretion scenario and that
the accreted dust has already settled out and leaves behind an
atmosphere with the respective C/O ratio.

\subsection{Cloud formation model}\label{ss:cloud_formation_model}
Our cloud formation model describes the formation of clouds by
nucleation, subsequent growth by chemical surface reactions on-top of
the seeds, evaporation, gravitational settling, element conservation
and convective replenishment
(\citealt{woi2003,woi2004,hell2006,hell2008b}; for a summery see
\citealt{hell2013}). The effect of nucleation, growth \& evaporation
on the remaining elements in the gas phase is fully accounted for
(Eqs. 10 in \citealt{hell2008b}).  The surface growth of a diversity
of materials causes the cloud particles to grow to $\mu$m-sized
particles of a mixed composition of those solids taken into
account. Cloud particles size and material composition change with
height in the atmosphere.

We have extended our model to allow a simultaneous treatment of
multiple nucleation species (seed formation) and to allow for the
formation and growth of mixes of carbonaceous and silicate materials.
Please refer to the above references regarding the formula body of our cloud formation model.

\paragraph{Seed formation:}
The seed formation is described by classical nucleation theory which
has been modified for TiO$_2$ nucleation to take into account
knowledge about (TiO$_2$)$_{\rm N}$ cluster formation
(\citealt{jeong1999,jeong2000,lee2015a}). We follow the approach,
incl. all material constants, as described in \cite{lee2015a} for
TiO$_2$ seed formation.

The second seed formation species that we take into account is carbon.
We do not present any new development of carbon-nucleation but apply
the approach used for AGB star wind modeling
e.g. in \citealt{flei1992,woi2006}, but for the more complete gas-phase
chemistry described in \cite{bilger2013}.

In an oxygen-rich gas (C/O$=0.43$), we apply TiO$_2$ nucleation only
as all carbon will be blocked in CO or CH$_4$. In a clearly carbon
rich case (C/O$=1.1, 1.5, 3.0, 10.0$), we apply carbon nucleation only
as all oxygen will be locked up in CO. We demonstrate in
Sect.~\ref{ss:CtoO=1} the simultaneous treatment of TiO$_2$ and C
nucleation for the case C/O=1.0 where the required oxygen and carbon are
not blocked by CO, and the gas phase composition becomes substantially more demanding.

\paragraph{Surface growth:}
The growth of the cloud particle bulk is described by gas-surface
reactions between the gas and the cloud particle surface as described
in \cite{hell2006} and \cite{hell2008b}. As different materials can
become thermally stable at very similar temperatures, a mix of
material will grow.  The surface reactions that were used are
summarized in Tables~\ref{tab:Cgrowth} and~\ref{tab:Ogrowth}.  For
carbon-rich cases, we apply the set of surface reaction listed in
Tables~\ref{tab:Cgrowth}, and for oxygen-rich cases
Table~\ref{tab:Ogrowth}. The chemical transition case of equal carbon
and oxygen abundances (C/O=1.0) is represented by
Table~\ref{tab:Ogrowth} with the addition of C[s], SiC[s], TiC[s],
KCl[s] and MgS[s] from Tables~\ref{tab:Cgrowth}.  The vapor pressure
data used for the materials that were added to our cloud model (C[s],
MgS[s], TiC[s], SiC[s], KCl[s]) are given in Table~\ref{tab:V0ci}.

\begin{figure*}
\centering
 \includegraphics[scale=0.47]{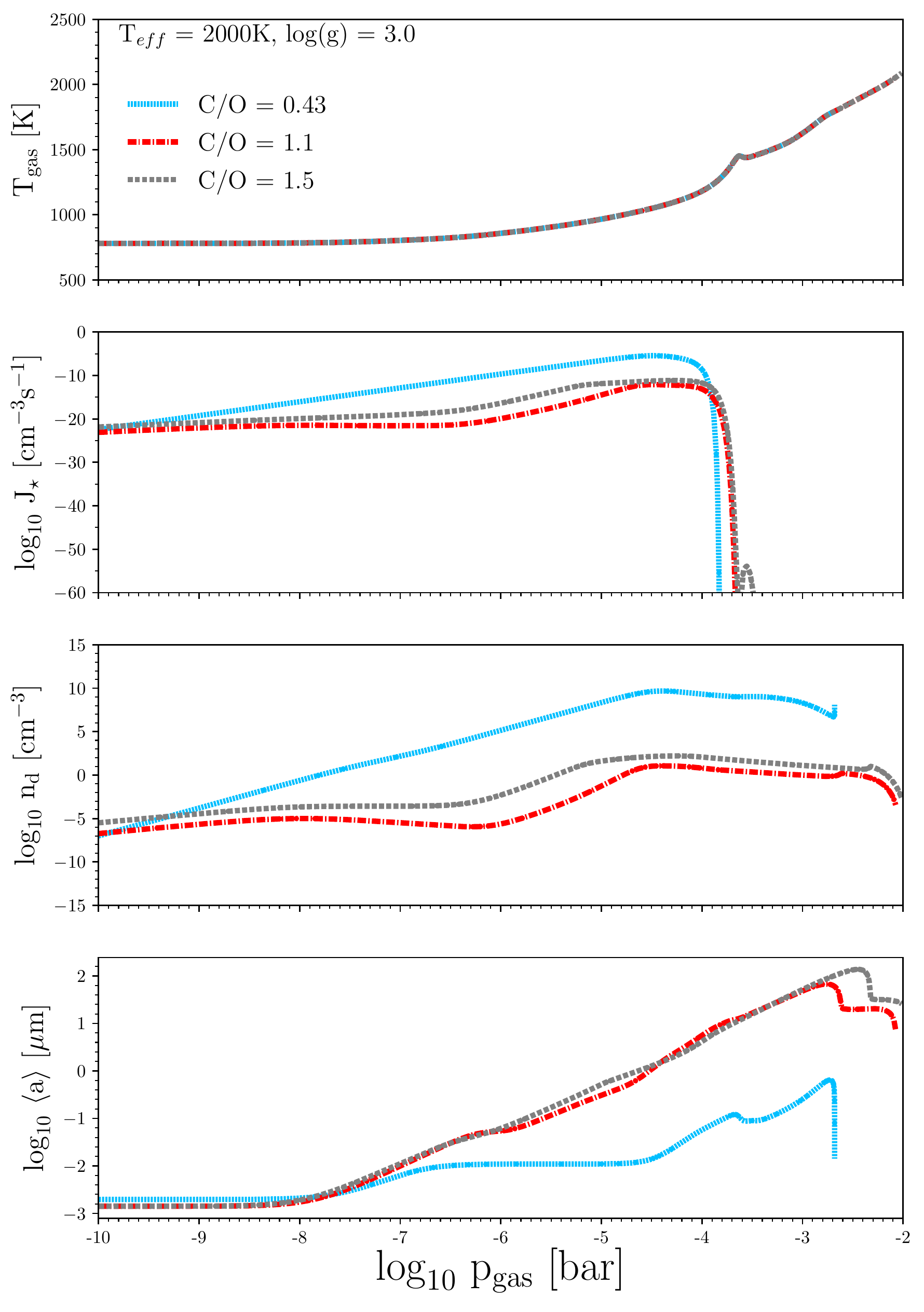}
 \includegraphics[scale=0.47]{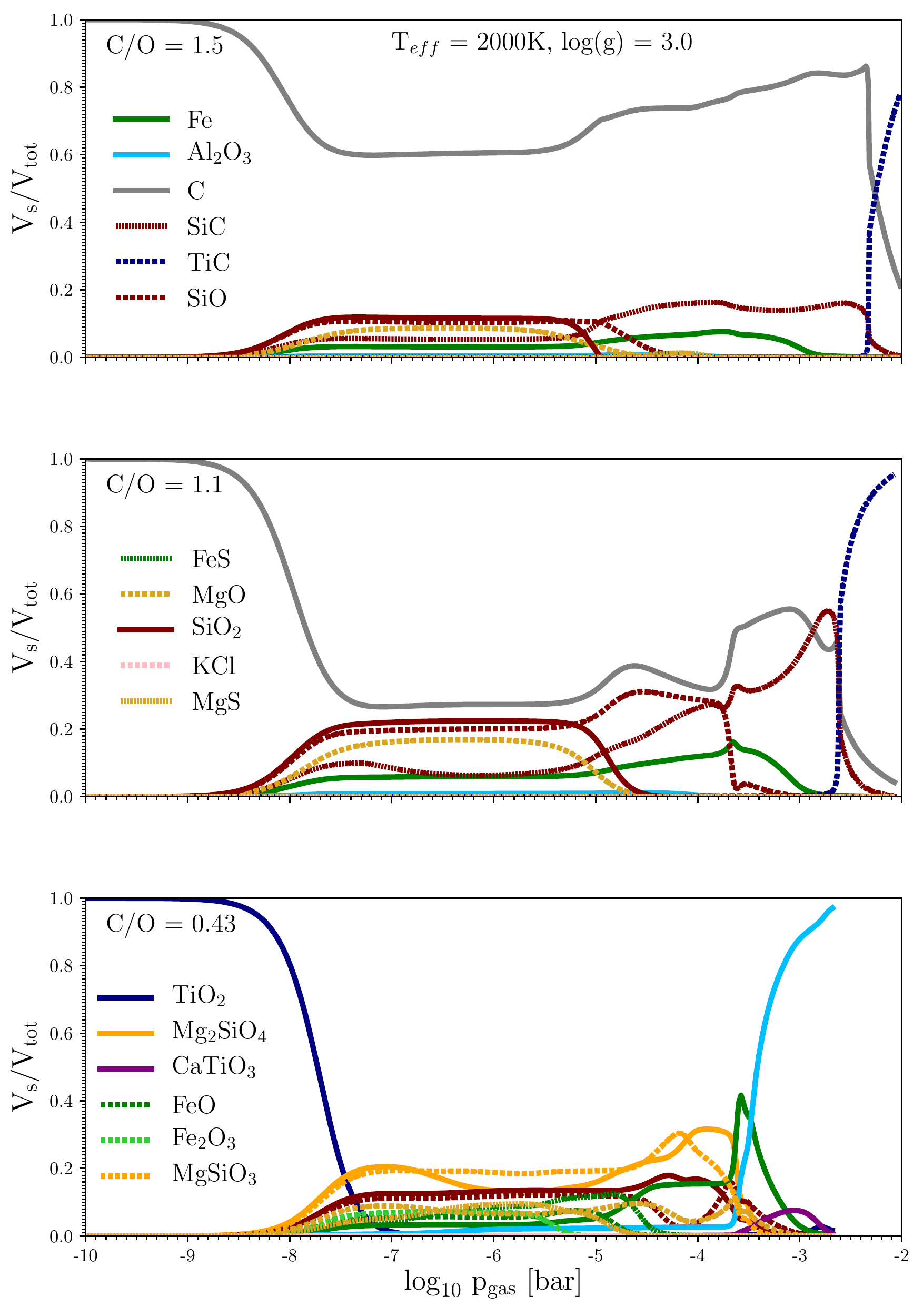}
 \caption{{\bf Left:} Cloud structures for changing C/O=0.43 (blue), 1.1 (red), 1.5 (grey)
   for a prescribed {\sc Drift-Phoenix} (T$_{\rm gas}$, p$_{\rm gas}$)
   structure for a giant gas planet with T$_{\rm eff}=2000$K,
   log(g)=3.0, initial solar element abundances. {\bf 1st panel:} input T$_{\rm gas}$ [K] and p$_{\rm gas}$ [bar], {\bf 2nd panel:} nucleation rate $\log J_*$ [cm$^{-3}$s$^{-1}$], {\bf 3rd  panel:} cloud particle number density $\log n_{\rm d}$  [cm$^{-3}$],  {\bf 4th  panel:}  mean cloud particle radius $\log \langle a \rangle$ [$\mu$m].\newline
{\bf Right:} Changing material composition, $V_{\rm s}/V_{\rm tot}$ [\%] (relative volume fractions of solid species $s$)  of the cloud particles with changing C/O ratio.}
\label{fig:CtoOchange}
\end{figure*}

\subsection{Drift-Phoenix model atmosphere}

{\sc Drift-Phoenix} (Dehn 2007, \citealt{hell2008}, \citealt{witte2009}) model
atmosphere simulations solve the classical 1D model atmosphere
problem coupled to a kinetic phase-non-equilibrium cloud formation
model. Each of the model atmospheres is determined by the effective
temperature (T$_{\rm eff}$ [K]), the surface gravity (log(g) (with g
in cm/s$^2$), and element abundances. The cloud's opacity is
calculated applying Mie and effective medium theory.

The 1D atmosphere models provide atmospheric properties such as the
local convective velocity, and the temperature-pressure (T$_{\rm gas}$
[K], p$_{\rm gas}$ [dyn/cm$^2$]) structure. The local temperature is
the result of the radiative transfer solution, the local gas pressure
of the hydrostatic equilibrium. For the lack of a
consistent solution as of yet, we apply the same oxygen-rich
atmosphere structure in all tests presented in this paper.


\section{Clouds in atmospheric environments of changing C/O ratio from  C/O$<$1 to C/O$>$1}\label{CtoOdiff}

\subsection{Changing cloud structures with changing C/O}\label{ss:Changing_cloud_structure}

The cloud structure of oxygen-rich giant gas planets and brown dwarfs
is well understood based on kinetic cloud modeling
(\citealt{hell2013,hell2014b}), but it still depends on the basic
material constants (e.g. vapor pressure, cluster data;
\citealt{fort2016}) and poses a challenge as part of atmosphere model
simulations (forward and retrieval) where it is most often
parameterised.  In a stationary scenario, haze-like small cloud
particles made of a rich mix of silicates and metal-oxides populate
the uppermost cloud layers. If they precipitated through the
atmosphere, the cloud particles change their size and the material
composition becomes dominated by Mg/Si/O materials with only small
inclusions from iron and other oxides. The innermost cloud part is
made of big particles with a mix of high-temperature condensates of
which one is always dominating (e.g. TiO$_2$[s] or Fe[s]). This is the
case for C/O=0.43 which is shown for reference in
Fig.~\ref{fig:CtoOchange_individual} (right). This cloud structure is
a refinement of the {\sc Drift-Phoenix} result as published in
\cite{witte2009}.

\begin{figure*}
\centering
 \includegraphics[scale=0.47]{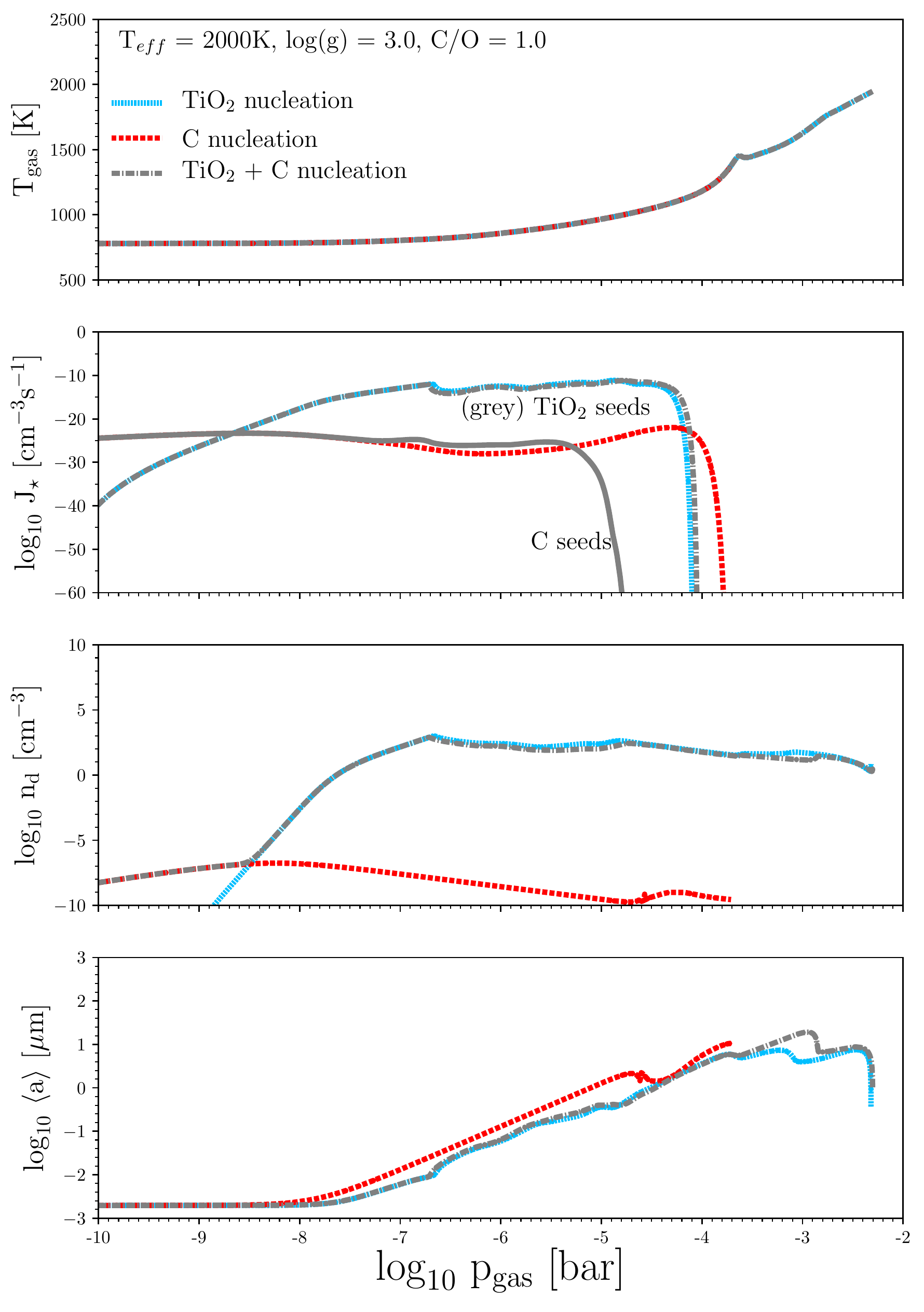}
 \includegraphics[scale=0.47]{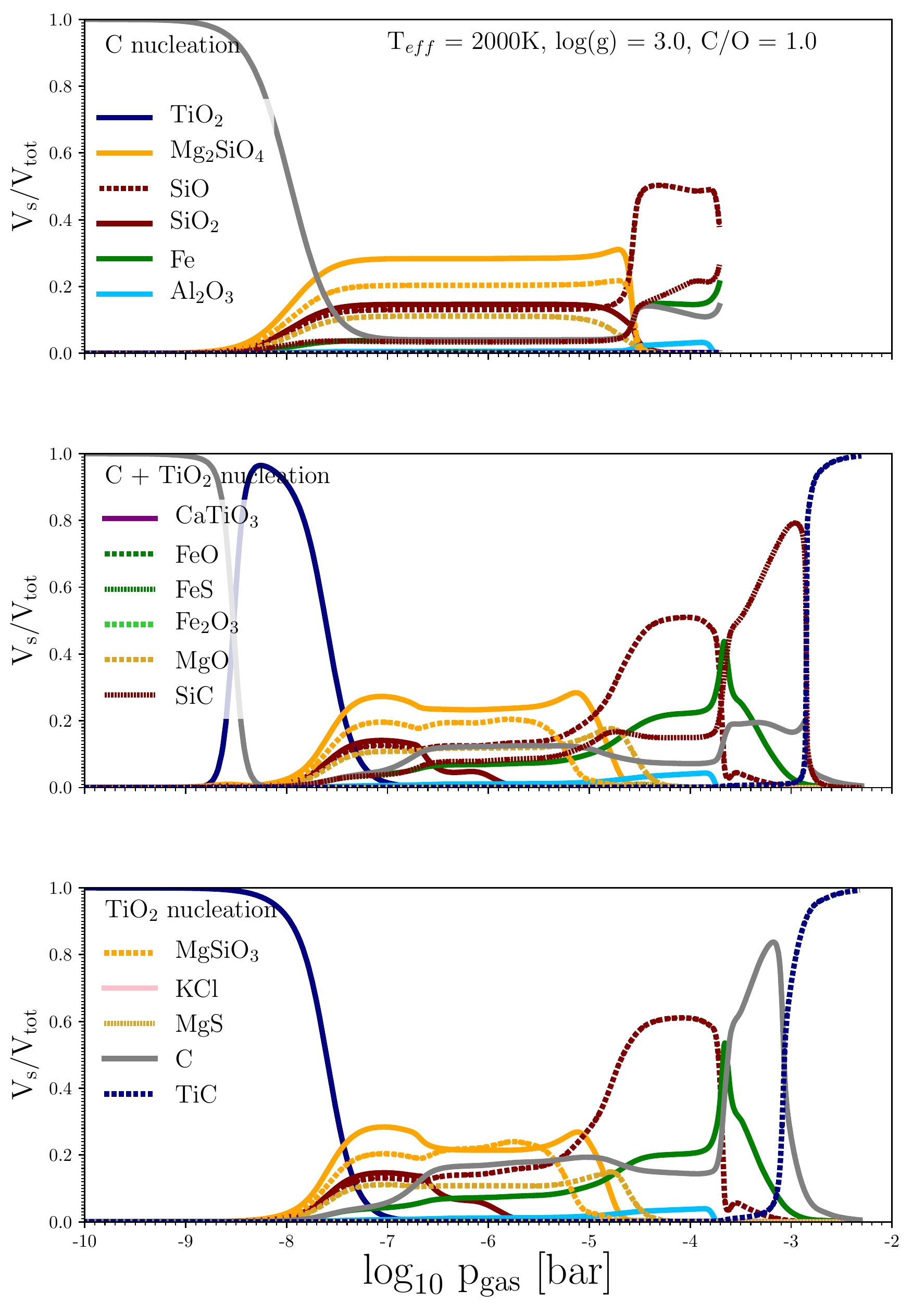}
 \caption{Clouds formation at C/O=1.0 for a prescribed {\sc
     Drift-Phoenix} (T$_{\rm gas}$, p$_{\rm gas}$) structure for a
   giant gas planet with T$_{\rm eff}=2000$K, log(g)=3.0, initial
   solar element abundances. Three cases are tested: i) TiO$_2$
   nucleation only (light blue) , ii) carbon nucleation only (red),
   iii) simultaneous carbon and TiO$_2$ nucleation (gray).  {\bf Left:} {\bf 1st
     panel:} input T$_{\rm gas}$ [K] and p$_{\rm gas}$ [bar], {\bf 2nd
     panel:} nucleation rate $\log J_*$ [cm$^{-3}$s$^{-1}$], {\bf 3rd
     panel:} cloud particle number density $\log n_{\rm d}$
   [cm$^{-3}$], {\bf 4th panel:} mean cloud particle radius $\log
   \langle a \rangle$ [$\mu$m].\newline 
    {\bf Right:} Changing material composition, $V_{\rm
     s}/V_{\rm tot}$ [\%] (relative volume fractions of solid species
   $s$) of the cloud particles for different nucleation species.}
\label{fig:CtoO=1}
\end{figure*}

\paragraph{Over-all cloud structure:} Figure~\ref{fig:CtoOchange} (left) demonstrates how the cloud structure 
changes if the initial C/O ratio (C/O$_0$) changes from an oxygen-rich (C/O$_0<$1) to a
carbon-rich (C/O$_0>$1) gas-composition.  Such a drastic change in
environmental chemistry can occur during disk evolution and also due
to the set-in of dust formation (\citealt{hell2014}). We present our
results for C/O$_0$=0.43 (blue), 1.1 (red) and 1.5 (gray).
Figure~\ref{fig:CtoOchange} (right) shows how the material composition
of the cloud particles (in units of $V_{\rm s}/V_{\rm tot}$ --
relative volume fractions of solid species $s$) change.

The results from the second panel in Fig.~\ref{fig:CtoOchange} (left)
show that the nucleation rate of TiO$_2$ (light blue line) in an
oxygen rich environment exceed that of C in a carbon rich gas (red
line) because carbon is initially considerably more abundant in the
gas-phase and therefore growth very quickly onto the newly formed
grain. This leads to a rapid decrease of carbon in the gas which
causes the nucleation process to stop. An increase in C/O ratio to
C/O$_0$=1.5, i.e. an increase of carbon, leads to a somewhat increased
seed formation rate (gray) compared to the C/O$_0$=1.1 case.

A larger nucleation rate will lead to more seeds forming and hence a
larger number density of cloud particles $n_{\rm d}$
(Fig.~\ref{fig:CtoOchange}, left, 3rd panel) for a given (T$_{\rm
  gas}$, p$_{\rm gas}$)-structure. The nucleation rates reach their
maximum at almost the same pressure level independent of the initial
C/O ratio. This might change if carbon-cloud and temperature structure
were consistently treated.  The cloud extends below the nucleation
regimes in all cases depicted in Fig.~\ref{fig:CtoOchange} (left,
compare 2nd \& 3rd panel) as the growing cloud particles
gravitationally settle into deeper atmospheric layers.

The bottom panel of Fig.~\ref{fig:CtoOchange} (left) shows that the
large number of seeds in the oxygen-rich case results in the smallest
average cloud particles sizes, as now more surface area is
available onto which a given number of gas species will grow. {\it As
  a result the cloud particles in a carbon rich environment are larger but less
  numerous.} It is interesting to note that the average particle size
of cloud particles in carbon-rich environments does not appear to differ
drastically between just carbon rich environments and more
concentrated carbon rich environments, despite the latter having a
higher number density by as as much as three orders of magnitude in
places. This suggests that there exists a much greater abundance of
potential condensible gas species in a rich carbon environment, such
that the increased number density is canceled out.

\paragraph{Cloud composition:}
Figure~\ref{fig:CtoOchange} (right) (see also
Fig.~\ref{fig:CtoOchange_individual} for comparison in more detail)
presents what materials the cloud particles are made of, how their
composition changes with height and how it changes with changing
C/O$_0$-ratio. The plotted values, $V_{\rm s}/V_{\rm tot}$, are the
relative volume fractions of the solid $s$ and are calculated from
Eqs.(3) in \cite{hell2008b}. Of particular interest is the question if
cloud particle in a carbon-rich atmosphere will remain homogeneous as
commonly assumed for AGB star wind modeling
(\citealt{flei1992,mat2010,wit2016}), or if they will be composed of
some mix of materials similar to the chemically more complex
oxygen-rich case.

In a heavily carbon rich atmosphere (here: C/O$_0$=1.5), the dominant
dust component is C[s], with it making up at least 60\% of the volume
except for at the deepest atmospheric layers where TiC[s] becomes more
prevalent. The top panel in Fig.~\ref{fig:CtoOchange} (right) also
shows the emergence of SiC[s] in the inner atmosphere, where it
becomes the second most abundant material between 10$^{-5}$ bars and
roughly 10$^{-3.5}$ bars. Other significant components of the dust
phase in the mid atmosphere are SiO[s], SiO$_2$[s] and MgO[s] reaching
a volume fraction of $\approx$30\% . Fe[s] appears in small amounts
that increase somewhat when SiO[s], SiO$_2$[s] and MgO[s] have
evaporated. \\ In comparison, a carbon rich atmosphere at a lower C/O
ratio (here: C/O$_0$=1.1), carbon, while still the most abundant dust
component for most of the cloud layer, comprises as little as 30\% of
the dust volume once other materials begin to condense onto the
cloud particles. The proportional increase of oxygen aids the condensation of
oxygen bearing dust species, such as SiO[s], SiO$_2$[s] and MgO[s] in
the mid cloud layer. SiO[s] exists for roughly an extra bar of
pressure in depth compared to C/O$_0$=1.5. The abundance of SiC[s]
also increases to a maximum of $\approx 50\%$ of the dust volume at
its peak, compared to just 20\% for a C/O$_0$ =1.5. Also Fe[s] has
increased its volume fraction to as much as 10$\ldots$ 15\%.\\ In an
oxygen rich environment (here: C/O$_0$=0.43) the seed particle TiO$_2$
is the prevalent species only during the nucleation period, in the
upper-most cloud layers where dust growth does not occur efficiently. Once
condensation onto the seed particles begins the dust volume instead
comprises of a mixture of Mg$_2$SiO$_4$[s], MgSiO$_3$[s] and many other silicate and oxide species.

To summarize; the seed particle material comprises a much more
significant fraction of the total dust volume in carbon rich cases,
due to the higher elemental abundances of C compared to Ti. In both
environments, silicon species seem to form a large portion of the
dust, bonded to Mg and multiple O in the oxygen rich environments, and
simply O or O$_2$ in the carbon rich.

\subsubsection{The case of C/O$_0$=1.0}\label{ss:CtoO=1}
Disk evolution seems to suggest that a chemical situation where
C/O$\approx$1 is not uncommon. ProDiMo\footnote{Radiation
    thermo-chemical models of protoplanetary disks}
  (\citealt{woi2009,woi2016}) chemical disk model results presented in
  \cite{hell2014} suggest that at 10Myrs a C/O$\approx$ 1.0 range
  appears between 1$\,\ldots\,10$AU. While the exact number will depend
  on the model set up (and on the chemical rates applied), the
  C/O$\approx 1.0$-range may well be a common features during certain
  evolutionary disk stages, or even indicative of it. Assuming that the
  disk dust has settled onto/into the planets core and that the planet
  is cold enough that no outgasing occurs, the planet is then left
  with a C/O$\approx$1.0 atmosphere.  Hence, an evolving planet will
  have to go through this stage of evolution at some point in its life
  time.  Here we use our cloud formation model to study the cloud
  details that occur should a planet be effected by the external disk
  C/O$\approx$1 or achieve else-wise C/O$\approx$1.

Modeling cloud formation in an atmosphere where carbon and oxygen
appear in the same amounts and are locked up on CO is challenging
because non of the well-studied cases (oxygen-rich and carbon-rich)
can be used as guide. While it is not obvious which nucleation species
to use (with the added challenge of the availability of material
data), we use the case to perform some tests on the effect of the
nucleation species on a potentially forming cloud. To do so, we perform
three test: i) TiO$_2$ nucleation only , ii) carbon nucleation only,
iii) simultaneous carbon and TiO$_2$ nucleation. All three test
calculations use the same set of surface growth reactions in
Table~\ref{tab:Ogrowth} with the addition of C[s], SiC[s], TiC[s],
KCl[s] and MgS[s] from Tables~\ref{tab:Cgrowth}. Hence, the only
difference is how we treat the formation of seed particles.

Figure~\ref{fig:CtoO=1} shows how the different nucleation species
affect the over-all cloud structure. The pure carbon-nucleation is the
least efficient seed formation process in the case C/O=1.0, pure
TiO$_2$ nucleation is the most efficient. The result is that pure
carbon-seeds would lead to very few cloud particles
(Fig.~\ref{fig:CtoO=1}, left, 3rd panel, red dashed lines) as it
occurs with 10 order of magnitudes less efficiency (2nd panel, red
dashed lines). It might, however, be unexpected that the mean cloud
particle radii (Fig.~\ref{fig:CtoO=1}, left, 4th panel) differ by not
too much.  Figure~\ref{fig:CtoO=1} (right) demonstrate the material
mix that can be expected for cloud particles forming from a gas of
initially C/O=1. While the upper cloud deck is dominated by the seed
forming species material, the remaining cloud where the surface growth
causes the largest increase in grain size, is rather similar between
the three cases shown in Fig.~\ref{fig:CtoO=1}. Given that the
nucleation process determines the whole cloud structure we suggest to
use carbon- and TiO$_2$-nucleation in order to optimally represent the
cloud structure in an atmospheres where C/O$\approx$1. The details of
the material composition will be discussed as part of our study of
varying C/O in Sect.\ref{ss:global_changes}.

\begin{figure*}
\centering
 \includegraphics[scale=0.47]{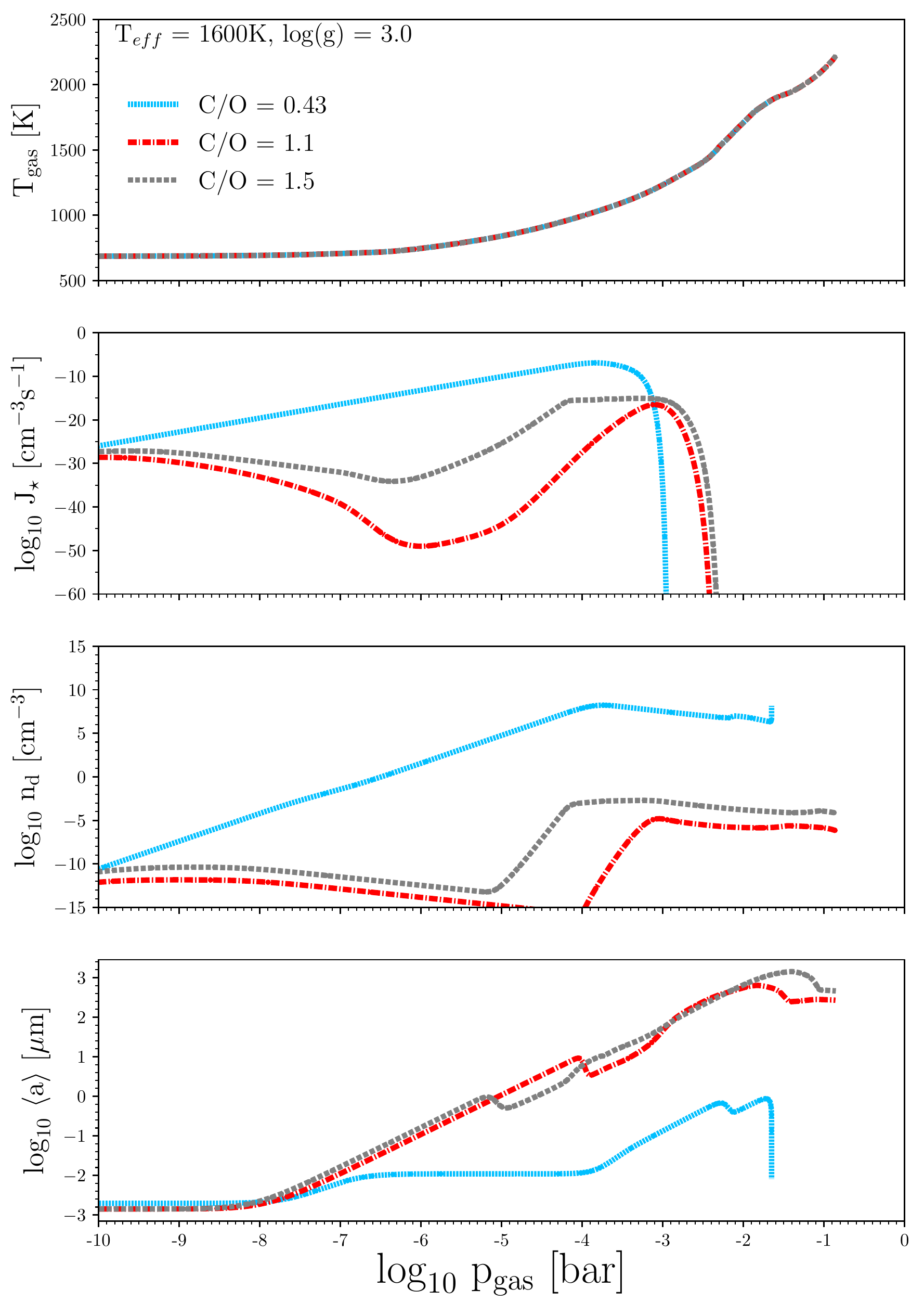}
 \includegraphics[scale=0.47]{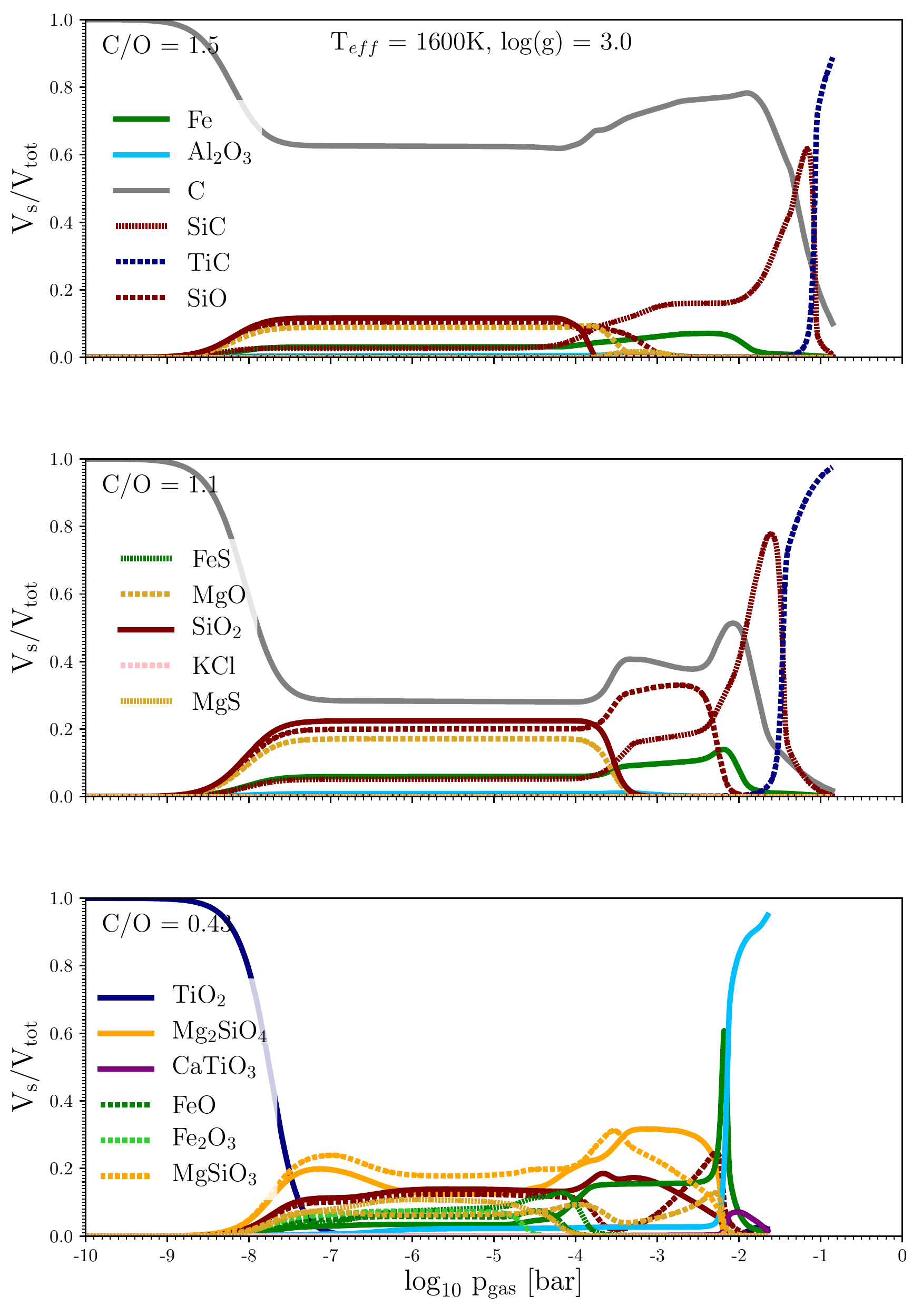}
 \caption{ Cloud structures for changing C/O=0.43 (blue), 1.1 (red), 1.5 (gray)  for a prescribed  {\sc Drift-Phoenix} (T$_{\rm gas}$, p$_{\rm gas}$)
   structure for a giant gas planet with T$_{\rm eff}=1600$K,
   log(g)=3.0, initial solar element abundances.  {\bf Left:} {\bf 1st panel:} input T$_{\rm gas}$ [K] and p$_{\rm gas}$ [bar], {\bf 2nd panel:} nucleation rate $\log J_*$ [cm$^{-3}$s$^{-1}$], {\bf 3rd  panel:} cloud particle number density $\log n_{\rm d}$  [cm$^{-3}$],  {\bf 4th  panel:}  mean cloud particle radius $\log \langle a \rangle$ [$\mu$m].\newline
{\bf Right:} Changing material composition, $V_{\rm s}/V_{\rm tot}$ [\%] (relative volume fractions of solid species $s$)  of the cloud particles for different nucleation species.}
\label{fig:CtoO=1_16}
\end{figure*}

\subsection{Decreasing T$_{\rm eff}$}
Given the strong temperature dependence of the cloud formation
processes, and in particular of the seed formation (see steep gradient
of J$_*$ in e.g. Figs.~\ref{fig:CtoOchange}, \ref{fig:CtoO=1}), we
examine the impact of decreasing T$_{\rm eff}$ of the atmosphere from
2000K to 1600K on the cloud structure and composition. The surface
gravity remains the same at log(g) = 3.0, and initial solar element
abundances for C/O$_0$=0.43, 1.1, 1.5 are studied.

\paragraph{Cloud Structure:}
Figure~\ref{fig:CtoO=1_16} shows for T$_{\rm eff}$=1600K (log(g)=3.0)
the changes in the cloud structure moving from the oxygen rich
environment (C/O$_0$ = 0.43) to carbon rich cases (C/O$_0$ = 1.1, 1.5).
As for the previous case, the nucleation rate of TiO$_2$ (second
panel, blue) in the oxygen rich case is greater than that of the
carbon nucleation in either carbon rich atmosphere. Compared to
T$_{\rm eff}$ = 2000K (Fig.~\ref{fig:CtoOchange}) the nucleation rate
is lower for all C/O ratios, but is particularly pronounced in mid
cloud layer (10$^{-7}$ to 10$^{-5}$ bars) in the carbon rich
environments where a dip in the nucleation rates of carbon seeds is
observed. The nucleation rate is of the order of 10 (C/O$_0$ = 1.5) to
20 (C/O$_0$ = 1.1) magnitudes smaller. Nucleation of seed particles
also continues for roughly a bar in depth deeper for all C/O regimes,
compared with T$_{\rm eff}$ = 2000K.

The lower nucleation rates have a dramatic impact on the number
density of cloud particles, particularly in the carbon rich cases. For the
oxygen rich environment, there is a four order decrease in magnitude
in the number density of cloud particles at the top of the cloud level at
T$_{\rm eff}$ = 1600K compared to T$_{\rm eff}$ = 2000K, however, this
difference becomes negligible deeper into the cloud structure. 
The number density increases at 10$^{-5}$ bars (in the present model)  for C/O$_0$ =
1.5 and 10$^{-4}$ bars for C/O$_0$ = 1.1 due to increasing nucleation
rates, which is more sharply pronounced than the slight increase
present for T$_{\rm eff}$ = 2000K. The minor decrease in number
density in the carbon rich regimes between 10$^{-8}$ bars and
10$^{-5}$/10$^{-4}$ bars is due to greater gravitational
settling, caused by the increased grain size, causing more cloud particles to rain out to lower cloud levels
than new seeds can replace.

As is expected, the average grain size (4th panel,
Figs.~\ref{fig:CtoOchange}, ~\ref{fig:CtoO=1_16}) remains much smaller
in the oxygen rich environments, where there is extra competition for
condensible material between the more numerous cloud particles, than for carbon
rich environments. Additionally, given the vastly lower number density
of cloud particles in an atmosphere with T$_{\rm eff}$ = 1600K, the cloud particles grow
to much larger radii than in the T$_{\rm eff}$ = 2000K case. At
the lower cloud levels, the average grain size is an order of
magnitude greater than for T$_{\rm eff}$ = 2000K. As in the hotter
environment however, the average grain size remains largely
independent of the degree of carbon abundance.

\begin{figure*}
 \hspace*{-0.3cm} \includegraphics[scale=0.55]{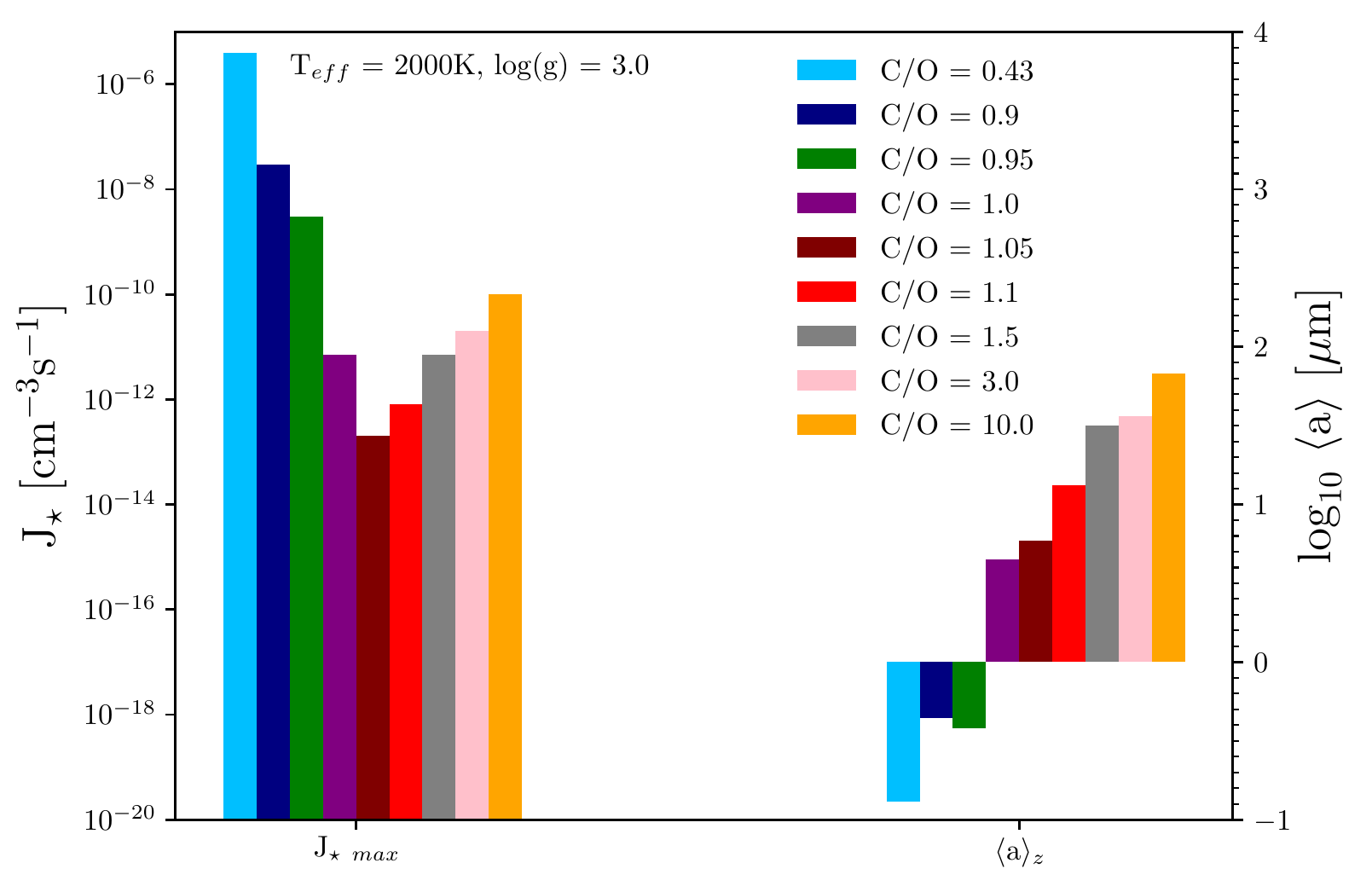}
 \includegraphics[scale=0.55]{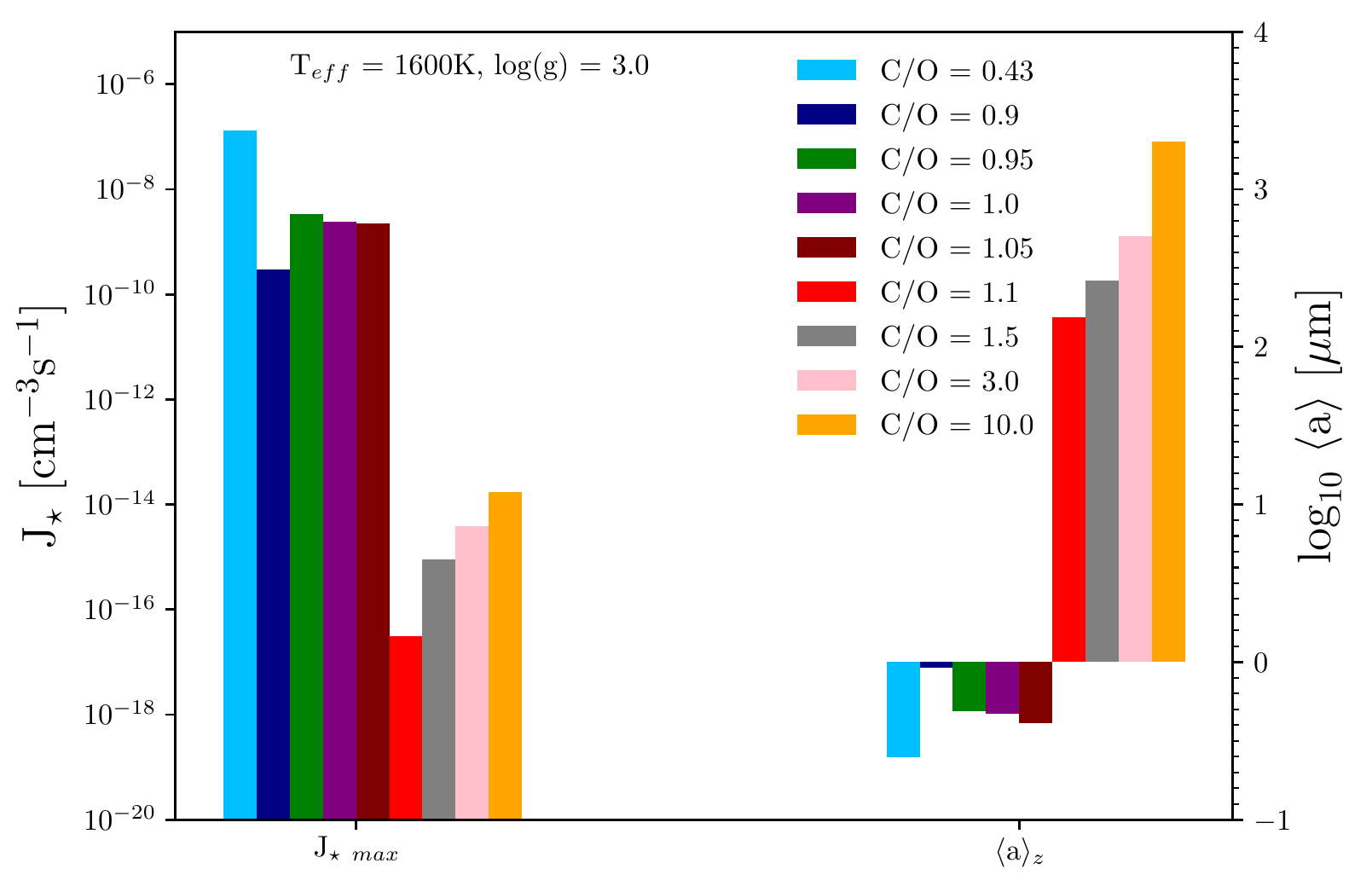}
 \caption{Global changes of cloud structure: Changes in maximum
   nucleation rate throughout cloud layer (left cluster of bars) and
   average mean grain size throughout cloud layer (right cluster of
   bars). Data taken from range of C/O$_0$ ratios from 0.43 to
   10.0.\newline {\bf Left} T$_{\rm eff}$ = 2000K and log(g) =3.0.
   {\bf Right} T$_{\rm eff}$ = 1600K and log(g) =3.0.}
\label{GlobalStructure}
\end{figure*} 

\paragraph{Cloud Composition:}
Figure~\ref{fig:CtoO=1_16} shows the composition of the cloud particles
throughout the cloud level. There is no change at the top of the cloud
level with all the dust comprising of seed particle material. As in
the T$_{\rm eff}$ = 2000K atmosphere there is a more significant
decrease in the carbon fraction of the dust for the less carbon rich
case. The condensation of other growth species again begins roughly
10$^{-8}$ bars with silicate and magnesium compounds comprising the
largest non carbon fractions of the dust (SiO[s], SiO$_2$[s] and MgO[s] for C/O$_0$
$>$ 1; Mg$_2$SiO$_4$[s], MgSiO$_3$[s] and SiO$_2$[s] for C/O$_0$ $<$ 1).

The composition of the dust in the bulk of the cloud structure, from
10$^{-8}$ bars to 10$^{-3}$ bars remain largely unchanged upon
altering T$_{\rm eff}$. The analysis of the differences between a
carbon rich cloud composition and oxygen rich cloud composition
presented in Section \ref{ss:Changing_cloud_structure} hold in this
region as well.

Changes in the cloud composition appear deep within the cloud
layer. In the oxygen rich environment the is a larger spike in the
volume fraction of Fe[s] prior to the sharp increase of Al$_2$O$_3$[s]
which then becomes the dominant fraction. In the carbon rich
environments SiC[s] becomes a major constituent of the dust volume,
roughly 70\% for C/O$_0$ = 1.5 and 80\% at C/O$_0$ = 1.1. This
compares to a maximum of roughly 15\% for C/O$_0$ = 1.5 and 60\% for
C/O$_0$ = 1.1 for T$_{\rm eff}$ = 2000K. The SiC[s] in the dust replaces
pure carbon suggesting that for cooler effective temperatures the deep
cloud layers become richer in Si-binding species.

  \begin{table}
 \label{tab:CloudStructure} 
\caption{Numerical comparison of Cloud structure over range of C/O ratios. Top of Cloud Layer corresponds to $1.43\cdot10^{-12}$ bars, middle of cloud layer to 10$^{-7}$ bars and bottom of cloud layer to between $1.37\cdot10^{-3}$ bars and $8.64\cdot10^{-2}$ bars, depending on the C/O ratio}
\centering
\begin{tabular}{p{1.5cm}p{1.3cm}|p{0.6cm}p{1.6cm}p{2.0cm}}
                                    &                            &                   &  V$_{\rm s}$/V$_{\rm tot}$ &\\ 
                                    &                             &       top & middle & bottom     \\
                \hline
C/O$_0$                             &  0.43                      &        100\%                 & 1.2\% TiO$_2$        & 1.7\% TiO$_2$   \\
J$_{*,max}$                & $4\cdot10^{-6}$  &            TiO$_2$          & 20.5\% Mg$_2$SiO$_4$ & 97\% Al$_2$O$_3$ \\
$[$cm$^{-3}$ s$^{-1}]$ &                                &                     & 11.2\% SiO           & 0.4\% Fe         \\
 $\langle a\rangle_{avg}$                  &  0.13                       &                     & 12.6\% SiO$_2$       & 0.8\% CaTiO$_3$  \\
 $[\mu$m$]$                 &                                &                     & 19.3\% MgSiO$_3$     &                  \\
          &                                  &                                        & 6.1\% Al$_2$O$_3$    &                  \\
          &                                  &                                        & 5.5\% FeO            &                  \\
          &                                  &                                        & 4.5\% FeS            &                  \\
          &                                  &                                        & 7\% Fe$_2$O$_3$       &                   \\
          &                                  &                                        & 8.9\% MgO             &                  \\
          &                                  &                                        & 5.3\% MgS             &                  \\\hline
C/O$_0$               &      0.9              &                   100\% & 1.2\% TiO$_2$          & 5.2\% TiO$_2$    \\
J$_{*,max}$ &      $3\cdot10^{-8}$ &              TiO$_2$                          & 25.1\% Mg$_2$SiO$_4$   & 0.2\% SiO        \\
$[$cm$^{-3}$ s$^{-1}]$ &         &                                        & 11.5\% SiO            & 92\% Al$_2$O$_3$  \\
 $\langle a\rangle_{avg}$          &     0.44   &                                        & 12.9\% SiO$_2$        & 1\% Fe            \\
 $[\mu$m$]$        &             &                                        & 18.1\% MgSiO$_3$      & 1.5\% CaTiO$_3$   \\
          &                                  &                                        & 0.6\% Al$_2$O$_3$     &                    \\
          &                                  &                                        & 5.9\% FeO             &                    \\
          &                                  &                                       & 2\% FeS               &                    \\
          &                                  &                                       & 7.5\% Fe$_2$O$_3$      &                    \\
          &                                  &                                        & 9.6\% MgO              &                    \\
          &                                  &                                        & 2.1\% MgS              &                    \\\hline
C/O$_0$      &    0.95               &         100\%& 1.2\% TiO$_2$          & 4.6\% SiO          \\
 J$_{*,max}$    &       $3\cdot10^{-9}$  &       TiO$_2$                                 & 26.1\% Mg$_2$SiO$_4$   & 33.9\% Al$_2$O$_3$ \\
  $[$cm$^{-3}$ s$^{-1}]$        &            &                                        & 12\% SiO               & 58.7\% Fe         \\
 $\langle a\rangle_{avg}$     &     0.38                            &                                        & 13.5\% SiO$_2$         & 1.5\% CaTiO$_3$    \\
$[\mu$m$]$           &   &                                        & 18.7\% MgSiO$_3$       & 1.2\% MgO          \\
          &                                  &                                        & 6.3\% FeO              &                    \\
          &                                  &                                        & 6.8\% Fe$_2$O$_3$      &                     \\
          &                                  &                                        & 9.6\% MgO              &                     \\
          &                                  &                                        & 2.1\% MgS              &                     \\\hline 
 \end{tabular}
\end{table}        
\begin{table}
\label{tab:CloudStructure2} 
\caption{Numerical comparison of Cloud structure over range of C/O ratios. Top of Cloud Layer corresponds to $1.43\cdot10^{-12}$ bars, middle of cloud layer to 10$^{-7}$ bars and bottom of cloud layer to between $1.37\cdot10^{-3}$ bars and $8.64\cdot10^{-2}$ bars, depending on the C/O ratio}
\centering
\begin{tabular}{p{1.5cm}p{1.3cm}|p{0.6cm}p{2cm}p{1.6cm}}
                                    &                            &                   &  V$_{\rm s}$/V$_{\rm tot}$ &\\ 
                                    &                             &       top & middle & bottom     \\
                \hline
C/O$_0$      &    1.0                &      100\%        & 1.3\% TiO$_2$           & 0.5\% C             \\
 J$_{*,max}$          &  $7\cdot10^{-12}$ &  C     & 27.2\% Mg$_2$SiO$_4$    & 99.3\% TiC          \\
 $[$cm$^{-3}$ s$^{-1}]$          &                                  &                                        & 12.5\% SiO              & 0.1\% Fe            \\
 $\langle a\rangle_{avg}$            &    4.47      &                                        & 14.1\% SiO$_2$          &                     \\
 $[\mu$m$]$           &                                  &                                        & 3.8\% Fe                &                     \\
          &                                  &                                        & 19.5\% MgSiO$_3$        &                      \\
          &                                  &                                        & 10.9\% MgO              &                      \\
          &                                  &                                        & 5.6\% C                 &                      \\ 
          &                                  &                                        & 4\% SiC                 &                      \\\hline  
C/O$_0$      &  1.05                  &                     100\%    & 1.4\% TiO$_2$        & 4.3\% C      \\
 J$_{*,max}$   &  $2\cdot10^{-13}$          C                      &                                    & 25.5\% Mg$_2$SiO$_4$ & 0.3\% SiC    \\
 $[$cm$^{-3}$ s$^{-1}]$          &                                  &                                    & 11.8\% SiO           & 95.4\% TiC   \\
 $\langle a\rangle_{avg}$           &    5.88       &                                    & 3.8\% SiO$_2$        & 0.8\% Fe      \\
 $[\mu$m$]$        &                &                                    & 18.3\% MgSiO$_3$     &               \\
          &                                  &                                    & 10.1\% MgO           &               \\
          &                                  &                                    & 11.3\% C             &               \\
          &                                  &                                    & 4.2\% SiC            &               \\\hline  
C/O$_0$       &    1.1                &                 100\%   & 5.9\% Fe             & 4.3\% C        \\
J$_{*,max}$  &     $8\cdot10^{-13}$   & C                                    & 0.5\% Al$_2$O$_3$    & 0.3\% SiC       \\
 $[$cm$^{-3}$ s$^{-1}]$   &                 &                                    & 26.7\% C             & 95.4\% TiC      \\
a$_{avg}$  &      13.31     &                 & 8.3\% SiC            &                 \\
$[\mu$m$]$          &                                  &                                    & 19.7\% SiO           &                 \\ 
          &                                  &                                    & 16.3\% MgO           &                 \\
          &                                  &                                    & 22.1\% SiO$_2$       &                 \\\hline
C/O$_0$     &  1.5                           &                 100\%   & 3.1\% Fe             & 21.2\% C        \\
J$_{*,max}$           &     $7\cdot10^{-12}$    &   C  & 0.2\% Al$_2$O$_3$    & 0.5\% SiC        \\
$[$cm$^{-3}$ s$^{-1}]$          &                                  &                                    & 59.9\% C             & 78.3\% TiC       \\
 $\langle a\rangle_{avg}$         &      31.60    &                                    & 5.5\% SiC            &                  \\
 $[\mu$m$]$     &                   &                                    & 10.8\% SiO           &                   \\
          &                                  &                                    & 8.5\% MgO            &                   \\
          &                                  &                                    & 11.8\% SiO$_2$       &                   \\\hline 
C/O$_0$       &   3.0                &             100\% & 1.2\% Fe             & 96.8\% C          \\
J$_{*,max}$           &    $2\cdot10^{-11}$  & C                                   & 84.4\% C             & 3.1\% SiC         \\
 $[$cm$^{-3}$ s$^{-1}]$         &         &                                    & 2.2\% SiC            &                   \\
 $\langle a\rangle_{avg}$               &    36.56      &                                    & 4.1\% SiO            &                    \\
 $[\mu$m$]$          &                                  &                                    & 3.3\% MgO            &                   \\
          &                                  &                                    & 4.5\% SiO$_2$        &                    \\\hline
C/O$_0$       &   10.0                                         &     100\%    & 0.3\% Fe             & 99.5\% C           \\
 J$_{*,max}$         &      $1\cdot10^{-10}$  &    C                                & 96\% C               & 0.4\% SiC          \\
 $[$cm$^{-3}$ s$^{-1}]$          &                                  &                                    & 0.6\% SiC            &                    \\
  $\langle a\rangle_{avg}$        &         67.38                         &                                    & 0.9\% SiO            &                    \\
 $[\mu$m$]$          &                                  &                                    & 0.9\% MgO            &                    \\
          &                                  &                                    & 1.1\% SiO$_2$        &                     \\\hline
\end{tabular}
\end{table}

\subsection{Global changes of cloud properties with changing C/O$_0$ , incl. the Extreme}\label{ss:global_changes}

We demonstrate how the globally changing C/O ratio effects global
cloud properties.  This may be envisioned as an evolutionary C/O
sequence from a typical oxygen-rich to a strongly carbon-rich
environment. Two scenarios maybe envisison for such evolutionary
  C/O sequence: a) an AGB star enriching the planet when it evolved
  away from the main sequence, and b) a planet encountering chemically
  different acretion environemnts in a planet-forming disk.  In this
first study, we wish to derive global cloud trends for changing C/O
ratios. We therefore also consider extreme values (C/O$_0$=3.0 and
10.0) which are guided by observations of carbon-rich Pop II stars
(e.g. Table 1 in \citealt{mas2016} and references therein).  The
results are summarizes in Table~\ref{tab:CloudStructure} and in
Fig.~\ref{GlobalStructure}.

Three C/O-regimes are considered: the oxygen rich regime
(C/O$_0$ = 0.43 and 0.9), the transition regime (C/O$_0$ = 0.95, 1.0
and 1.05) and the carbon rich regime (C/O$_0$ = 1.1, 1.5, 3.0 and
10.0). The oxygen and carbon rich cases were treated with a single
seed species (respectively TiO$_2$ and C) as described in
Sect.~\ref{ss:cloud_formation_model}. The transition region was
treated with both seed species and an increased selection of growth
species as described in Sect.~\ref{ss:CtoO=1}.

We utilize the following properties to provide a global
characterization of the cloud: the maximum nucleation rate, $J_{\rm *,
  max}$ $[$cm$^{-3}$ s$^{-1}]$, the average mean cloud particle size
$\langle a\rangle_{\rm avg}$ [$\mu$m] (mean cloud particle radius,
$\langle a\rangle$, averaged over whole vertical, 1D cloud extension), and three
characteristic values for the cloud particles material composition
(top, middle and bottom of cloud).

\paragraph{Maximum nucleation rate:}
The maximum nucleation rate, $J_{\rm *, max}$, within the cloud structure (left cluster
of bars in Fig.~\ref{GlobalStructure}) decreases as the initial C/O ratio
approaches unity from the oxygen rich side. This is due to
the decrease in oxygen not locked up in CO resulting in a gas phase
abundance of TiO$_2$ with which to form seed particles. The maximum
nucleation rate then increases again as the initial C/O increase above unity,
again  due to the increased abundance of gas phase seed
material, in this case carbon. The maximum
nucleation rate can be  greater at unity than at C/O$_0$ = 1.05 due to the
greater availability of TiO$_2$ in the gas phase with which to form
seeds. TiO$_2$ is a more efficient seed forming species than C,
which results in higher maximum nucleation rates even if
the gas phase abundance of carbon is much greater (such as
 in the extreme C/O$_0$ = 10 compared with the slightly
oxygen rich environment of C/O$_0$ = 0.9). This trend is also present for T$_{\rm eff}=1600$K (Fig.~\ref{GlobalStructure} right) though less pronounced in the transition regime (C/O$_0$ = 0.95, 1.0 and 1.05).

\paragraph{Average mean cloud particle size:}
The average mean grain size,  $\langle a\rangle_{\rm avg}$,  across the whole cloud layer (right
cluster of bars on Fig.~\ref{GlobalStructure}) shows an increasing
trend with increasing global  C/O$_0$. From previous sections, it was anticipated that the cloud particles present in the carbon rich
atmosphere would be larger than those in the oxygen rich atmosphere,
due to the decreased competition for condensible growth material as a
result of less seed particles forming. The increase in the averaged mean
grain size,  $\langle a\rangle_{\rm avg}$,  as the initial C/O ratio increases in the oxygen-rich regime is also cause by lower number densities of seed particles. The increasing average mean particle size,  $\langle a\rangle_{\rm avg}$,  with increasing initial C/O ratio in the carbon-rich regime,
however, may not be explained by this reasoning as the nucleation rate
also increases as the initial C/O ratio increases. Therefore, 
 the increased $\langle a\rangle_{\rm avg}$ are instead the result of
increased availability of growth material, specifically with regards
to carbon as a growth species. This is a conjecture supported by the
high proportion of carbon present in the dust composition of cloud particles at
almost every part of the cloud layer, as presented in the top two
panels of Fig.~\ref{fig:CtoOchange} (right).

\paragraph{Dust Composition:} 
The upper level of the cloud layer for every initial C/O ratio is made of 
the respective seed material. For the transition regime (C/O$_0$ = 0.95, 1.0
and 1.05), the C/O ratio decides which seed condenses first with
C/O $<$ 1 producing TiO$_2$ seeds and C/O$>=$ 1 producing carbon
seeds. The cloud particle volume in the middle of the cloud
layer is made of numerous materials. In the oxygen rich and
transition regimes the largest fraction of the volume are
Mg$_2$SiO$_4$[s], MgSiO$_3$[s], SiO[s] and SiO$_2$[s]. The percentage of
Mg$_2$SiO$_4$[s] grows slightly as the C/O$_0$  increases with
the small fraction of Al$_2$O$_3$[s] present at C/O$_0$ = 0.43 becoming
negligible as the C/O$_0$ ratio increases. In contrast, the largest
components of the normal carbon rich environments (C/O$_0$ = 1.1 and 1.5)
replace the fractions of Mg$_2$SiO$_4$[s] and MgSiO$_3$[s] with larger
proportions of SiO[s], SiO$_2$[s] and MgO[s] as well C[s]. As the C/O$_0$ ratio
increases, the fraction of C[s] increases too at the expense of SiO[s],
SiO$_2$[s] and MgO[s]. This trend is continued to the extreme C/O$_0$ ratios of
3.0 and 10.0 where the dust volume at the middle of the cloud layer is
almost completely comprised of carbon. At the bottom of the cloud
layer more stable dust species such as Al$_2$O$_3$[s] dominate in the
oxygen rich regime, though with volume fractions that decrease as the
C/O$_0$ ratio increase with increasing fractions of Fe[s] becoming present,
most notably when C/O$_0$ = 0.95. If C/O$_0\geq 1$,  TiC[s] becomes
the major material of the cloud particles, but decreases as the initial C/O increases,
with C[s] becoming more prevalent, indeed dominating the volume
fraction at extreme initial C/O ratios (3.0, 10.0).

\begin{figure*}
\centering
 \includegraphics[scale=0.47]{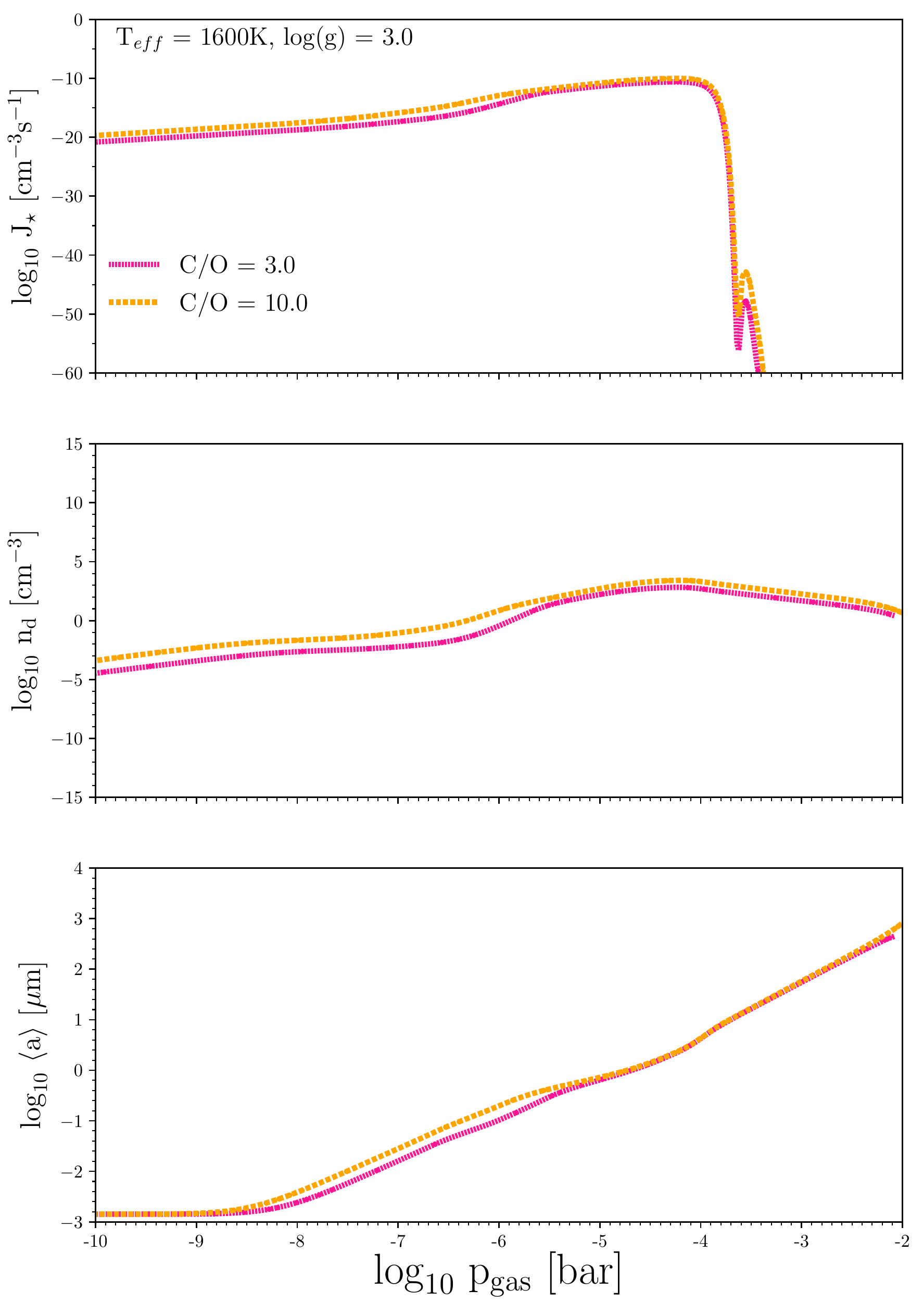}
 \includegraphics[scale=0.47]{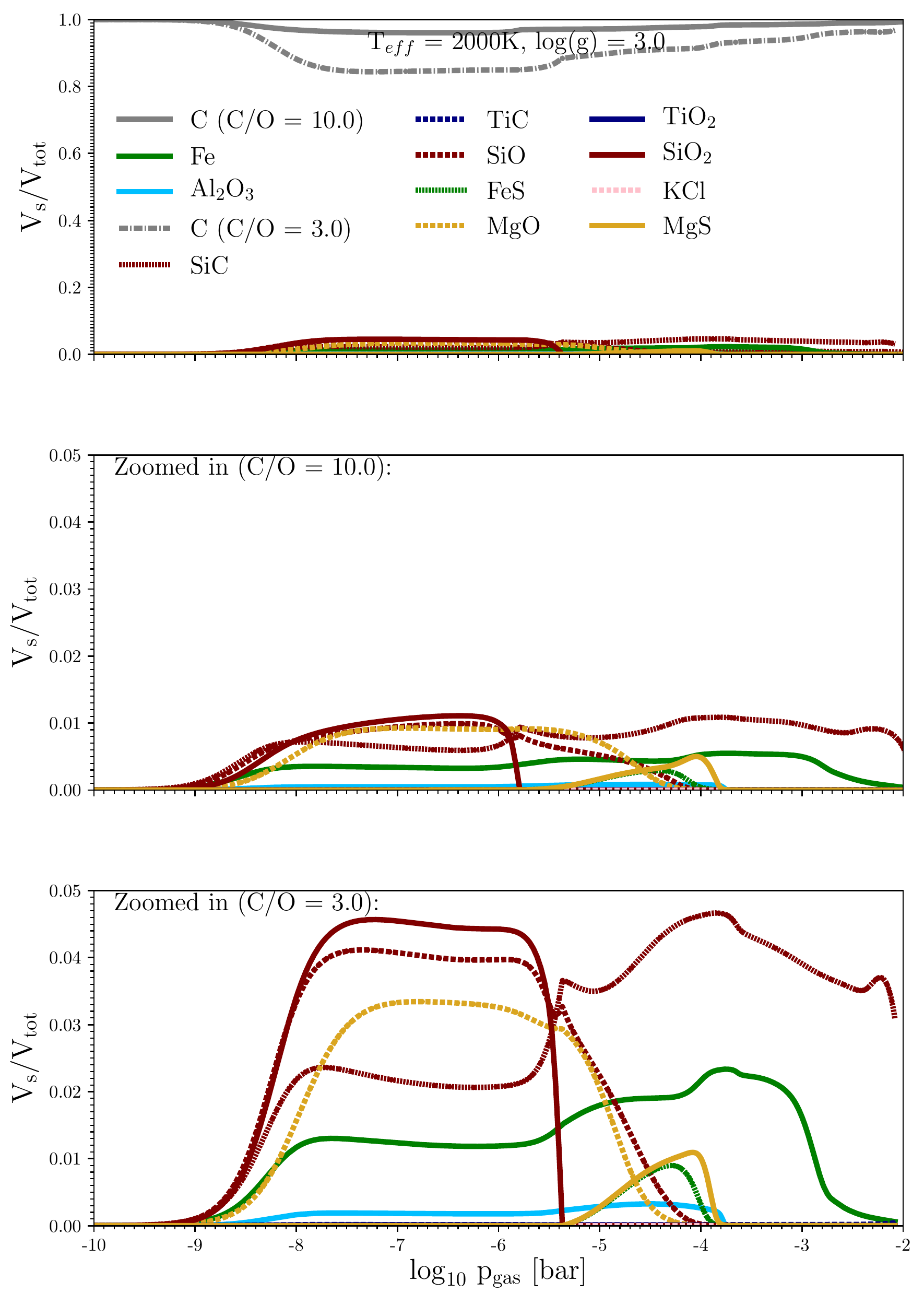}
 \caption{ {\bf Left:} Cloud structures for changing C/O=3.0 (pink), 10.0 (orange) for a prescribed  {\sc Drift-Phoenix} (T$_{\rm gas}$, p$_{\rm gas}$)    structure for a giant gas planet with T$_{\rm eff}=2000$K,
   log(g)=3.0, initial solar element abundances. {\bf 1st panel:} nucleation rate $\log J_*$ [cm$^{-3}$s$^{-1}$], {\bf 2nd  panel:} cloud particle number density $\log n_{\rm d}$  [cm$^{-3}$],  {\bf 3rd  panel:}  mean cloud particle radius $\log \langle a \rangle$ [$\mu$m].\newline
   {\bf Right:} {\bf 1st panel} Changing material composition, $V_{\rm s}/V_{\rm tot}$ [\%] (relative volume fractions of solid species $s$)  of the cloud particles for C/O ratios = 3.0 and 10.0.}
 \label{fig:PhysExtreme}
\end{figure*}

\subsection{Extreme C/O$_0$}\label{ss:extreme_C/O}

We briefly examine the cloud structure and cloud composition of
atmospheres extremely rich in carbon (C/O$_0$ = 3.0, 10.0). Only such
extremely carbon-rich cases produce cloud or dust particles that are
made almost entirely of carbon with only very little inclusions of
other materials. These results are not only of interest for planetary
atmosphere research but also for modeling the ISM enrichment
through AGB star winds and the enrichment of the early universe due to
SNe ejecta (e.g. \citealt{mas2016}). In particular the in situ
formation of carbon-rich planets in the early universe as by-product
of CEMP stars would occur at C/O$_0>10.0$ (Table 1 in
\citealt{mas2016}).

\paragraph{Cloud Structure:}
Figure~\ref{fig:PhysExtreme} illustrates the cloud structure in
extremely carbon rich environments. The general trend of the
nucleation rate (top panel), number density (second panel) and average
grain size (third panel) appear to be very similar to those presented
for C/O$_0$ = 1.1 and 1.5 in Fig.~\ref{fig:CtoOchange}, though somewhat larger. This is
expected for the nucleation rate (higher C/O ratio equates to a larger
abundance of potential seed material in gaseous C) and hence the
number density. The increased grain sizes is again caused by the increased abundance of growth
material (here: carbon). It is interesting to note that increasing the
C/O$_0$ ratio dramatically from 3.0 to 10.0 does not cause
dramatic changes in the cloud structure, a one order of magnitude in the
grain number density in the the upper half of the cloud layer being
the most notable difference.

\paragraph{Dust Composition:}
A dramatic effect of the increased initial C/O ratio from 3.0 to 10.0
is seen however, in Fig.~\ref{fig:PhysExtreme} (right). This figure
is demonstrating the dust composition in the extreme carbon rich
environments. At C/O$_0$ = 10.0, no less than roughly 95\% of the dust
volume is carbon at any point in the cloud layer, with SiO[s],
SiO$_2$[s], SiC[s] and MgO[s] making up only 1\% of the remaining
fraction each in the lower and middle cloud (2nd panel of
Fig.~\ref{fig:PhysExtreme} r.h.s.; this is a zoom-in of the 1st
panel). In contrast, when the C/O$_0$ =3.0 the carbon fraction
drops to roughly 85\% of the total volume in the upper middle portion
of the cloud, with Panel 3 showing SiO$_2$[s] and SiO[s] fractions around 4
times greater than present in C/O$_0$ = 10.0 and a MgO[s] fraction roughly 3
times as large. In the lower portions of the cloud, these fractions
evaporate and are replaced by a SiC[s] and Fe[s] with
each roughly a 4 times greater fraction than their equivalent fractions in the
C/O$_0$ = 10.0 case.


\begin{figure*}
\centering
\includegraphics[scale=0.47]{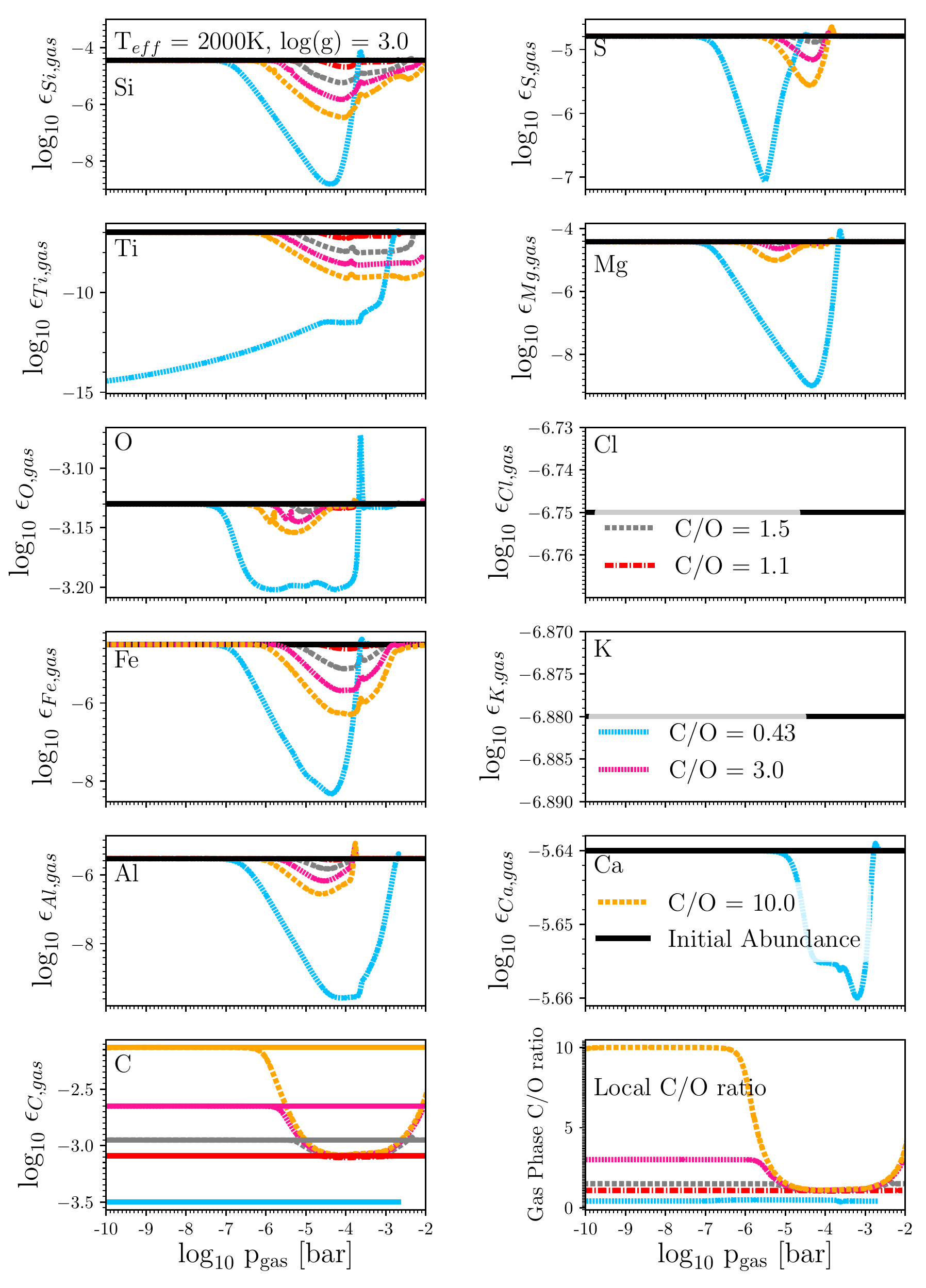}
 \includegraphics[scale=0.47]{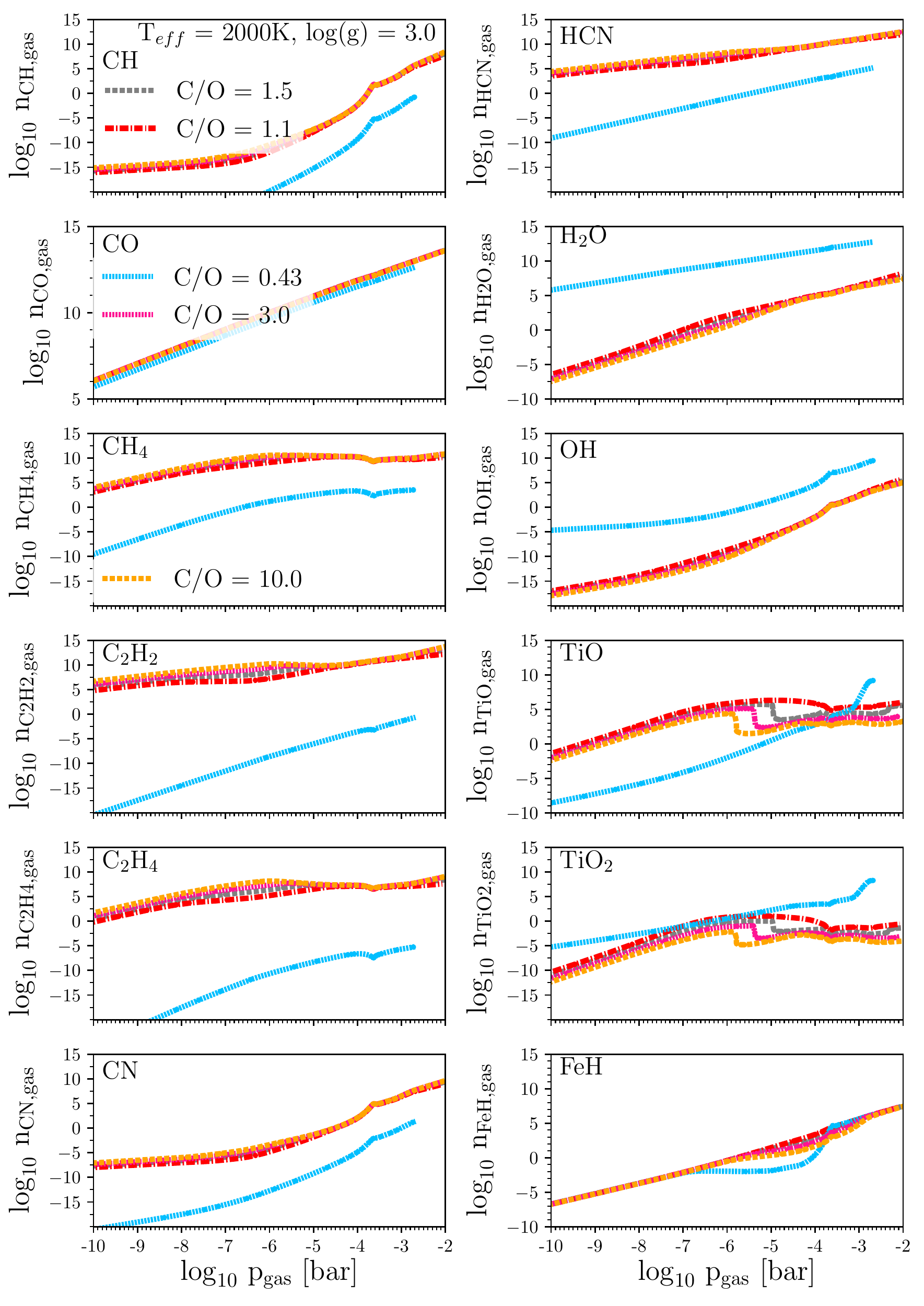}
\caption{Remaining gas-phase chemistry after cloud formation: {\bf Left:} depleted gas element abundances  after cloud formation. {\bf Right:} Molecular abundances of a selected number of species to demonstrate the differences between the oxygen-rich (blue) and the carbon-rich  cases. The underlaying atmosphere model is the same as in Fig.~\ref{fig:CtoOchange}  with T$_{\rm eff}$=2000K, log(g)=3.0 and initial solar abundances with the carbon adjusted according to the C/O ration listed.}
\label{fig:mol}
\end{figure*}

\section{Element depletion}\label{s:elm}

The element abundances, $\epsilon_{\rm x}$, determine the chemical
composition of the atmospheric gas from which the cloud particles
form. The cloud formation processes reduce the element abundances
through nucleation and surface growth or enrich them where evaporation
occur. We therefore study the effect of a changing C/O$_0$ ratio on
the element abundances remaining in the gas phase after the cloud
particles have formed. All element abundances are plotted in
comparison to the initial, solar element abundances (straight solid
line) with $\epsilon_{\rm C}$ adjusted according to the C/O$_0$
ratio. Figure~\ref{fig:mol} (left, panels depict different ranges)
show how differently the elements that participate in cloud formation
are depleted if the C/O$_0$ = 0.43, 1.0, 1.1, 1.5.

The by orders of magnitude largest element depletion for the C/O$_0$
ratios studied here occurs for C/O$_0$=0.43, i.e. in the oxygen-rich
case for Si, S, Ti, Mg, O, Fe, Al, and Ca. This is supported by our
findings that a considerably larger number of cloud particles composes
a cloud in an oxygen-rich compared to a carbon-rich cloud atmosphere,
despite remaining smaller in size. The element depletion becomes
generally smaller when approaching C/O$_0$=1.0 (compare C/O$_0$=1.5,
1.1, 1.0), and increases again when the case is clearly carbon-rich
like e.g. for C/O$_0$=1.5. The larges carbon-depletion occurs for
C/O$_0$=1.5 as now the carbon contributes the largest volume fraction
until its thermal stability limit near to bottom of the
cloud. Somewhat more cloud particles form for C/O$_0$=1.5 compared to
C/O$_0$=1.1 which does affect the element depletion (see
Fig.~\ref{fig:CtoOchange}).

Different nucleation species affect the chemical feedback on the gas phase differently. Ti is the most depleted element in the oxygen-rich case, while carbon remains moderately abundant even at very high C/O$_0$. An interesting case occurs in the transition regime where C/O$_0\approx$1.0.
Here,  carbon-nucleation alone is insufficient in describing element depletion.
 Both, TiO$_2$ nucleation and the simultaneous consideration of carbon and TiO$_2$ (C + TiO$_2$) produce very similar results, only that the simultaneous C + TiO$_2$ nucleation leads to a somewhat stronger depletion of Ti, and to less depletion of Fe, Al, S, and Ca. These differences are, however, very small.

K and Cl are not depleted for non of the C/O$_0$ ratios considered. This is in agreement with that KCl material volume fraction $\approx 0$ in Fig.~\ref{fig:CtoOchange} and ~\ref{fig:CtoO=1} (both r.h.s.).  KCl is therefore a super-inefficient growth species in the temperature regime represented by our T$_{\rm eff}$=2000K, log(g)=3.0 (oxygen-rich, initial solar element abundances) {\sc Drift-Phoenix} model atmosphere.

\subsection{The local C/O ratio}
The last row in Figures~\ref{fig:mol}  (left panel right column)  also shows the local, height-dependent C/O ratio for different global, initial C/O$_0$ ratios=0.43, 1.0, 1.1., 1.5., 3.0 and 10.0. These results  combine the depleted oxygen and the carbon abundances  effected by cloud formation. We find that the carbon-depletion in the carbon-rich cases makes the atmosphere locally somewhat more oxygen-rich by decreasing the locale C/O ratio. The largest change in the local C/O occurs for the highest value considered, C/O$_0$=10. All carbon-rich cases approach approximately the same local C/O$\approx 1.0$ in the atmospheric region with the strongest carbon depletion. This is also apparent from $\epsilon_{\rm C}$ in the same figure (left column). The evaporation of the carbon at the cloud bottom leads to a small increase of the C/O ratio above the initial value which is strongest for the highest C/O$_0$. As the oxygen-abundance decreases by cloud formation in oxygen-rich atmospheres, the local C/O ratio increases somewhat. This has been discussed in \cite{bilger2013} for different global parameters (T$_{\rm eff}$, log(g), [M/H]). \cite{hell2014} demonstrated that oxygen-depletion for a global C/O=0.99 can lead to a substantial increase in C/O locally.

\section{The abundances of  gas phase molecules}\label{sec:molabund}

Oxygen-rich gases (incl. CO, H$_2$O) and carbon-rich gases (incl. CO, CH$_4$) are dominated by different molecules (Fig.~\ref{fig:mol}). CO blocks the oxygen in carbon-rich gases and the carbon in oxygen-rich gases, unless the gas pressure is too high. In cool atmospheres with high densities, CH$_4$ becomes the most important carbon-binding species. Figure ~\ref{fig:mol} (right) shows the CO blocking works well in the case studied here.  

Figure ~\ref{fig:mol} further demonstrates the abundance of a number of selected molecules after cloud formation has reduced the element abundances as shown in Fig.~\ref{fig:mol}:  H$_2$O is the dominating oxygen-binding molecule regardless of the oxygen-depletion. 
TiO drops in abundance substantially and would produce a false-positive of a carbon-rich atmosphere due to the strong Ti depletion as result of cloud formation. The initial low Ti element abundances make TiO a good tracer for chemical effects on the atmospheric oxygen reservoir: TiO traces well the oxygen-depletion in the center of the cloud also in carbon-rich atmospheres (yellow to red lines),  and the oxygen-enrichment at the bottom of the cloud in the oxygen-rich case (blue lines). The TiO abundance returns to undepleted oxygen-rich values only when cloud particles evaporation has caused a considerable oxygen enrichment (see peak - blue line -  on $\epsilon_O$ on left of  Fig.~\ref{fig:mol}). FeH, as one example for metal-hydrate molecules, does reflect the Fe-depletion in the oxygen-rich case. Fe[s] contributes only with $1\,\ldots\,3\%$ to the particle volume in the carbon-rich cases, hence, causing only little effect on the FeH abundance. All other hydrates that bind elements , that participate into cloud formation, show a similar behavior. Metal hydrates can therefore appear relatively stronger in a spectrum of a cloud-forming atmosphere.

A carbon-rich atmosphere is dominated by more complex molecules like HCN, CH$_4$ and CO, with small carbo-hydrates like C$_2$H$_2$ and 
C$_2$H$_4$ increasing in abundance with increasing carbon-content.  Despite C$_2$H$_2$ being more abundant than CN, CN has a larger absorption cross  section and it would therefore be easier picked up from a spectrum (see Fig. 9 in \citealt{hell2000}).

\section{Concluding remarks}\label{s:sum}

We present the first cloud formation model that is applicable to 1D
atmosphere simulations of carbon-rich planetary objects. Carbon-rich
atmospheres of planets might indicate a particular chemical niche
during the chemical evolution of a protoplanetary
disk. Carbon-planets, however, were also suggested to form in the
early universe as by-product of Pop II stars. However, observing those
might only be possible of they e.g.  emerge as free-floating planets
nearby.

Clouds forming in carbon-rich atmosphere differ from their oxygen-rich
counterparts in that less cloud particles form which grow to larger
sizes and a large fraction of their material is made of
carbon. Similar to the oxygen-rich case, the cloud properties change
with height and the cloud particles are made of a mix of
materials. Cloud particles can only be nearly-homogeneous if the
initial C/O ratio is $\sim 10.0$. Such high C/O ratios are so far
predicted from low-metallicity theoretical AGB star models or for
carbon-enhanced metal-poor stars. 

When cloud particles rain (gravitational settle) into the deeper
atmosphere, they encounter increasing densities and increasing
temperatures. The increasing densities increase the thermal stability
of the materials and speed up the growth process due to an increase of
surface reactions. The temperature, however, will allow the material
to restructure into well-ordered matrices as pointed out
by \cite{hell2009}. Therefore, amorphous materials like carbon can turn into its
crystalline counterpart, i.e. into diamonds, which would cause drastic
change of optical properties during the cloud formation
process. Clouds on super-carbon-rich planets may therefore effect the
spectral appearance far less than in their oxygen-rich counterparts.

Carbon dust has been studied in great detail to understand the
enrichment of the ISM by stellar winds on the AGB which yield the
pre-courser for star and planet formation. \cite{and1999} demonstrate
that the extinction coefficient of amorphous carbon has no spectral
feature for $\lambda>0.3\mu$m (their Fig. 1), the exact value
depending somewhat on the data source used. Silicon carbide (SiC[s])
is another material that forms from AGB star winds which has been
identify from a broad emission feature $11\,\ldots\,11.5\mu$m
(Sect.~3.3 in \citealt{and1999}, \citealt{suh2000}). However, aromatic
3.3 $\mu$m PAH feature and the aliphatic sub-features in the range of
$3.4\,\ldots\,3.6 \mu$m were observed in the ISM and interpreted as a
mix of HAC and PAH absorption representing small dust grains
(\citealt{gad2013}). The transmission spectrum of the hot Jupiter
WASP-12b is almost completely flat. The optical slope appears
considerably shallower compared to HD\,189744b
(\citealt{sing2016}). While it is almost impossible that pure carbon
can condense in any oxygen-rich atmosphere based on an equilibrium
gas-phase chemistry composition, carbon and carbonaceous material
appears in abundance if C/O$>$1. It might therefore be thinkable that
either i) the atmosphere of WASP-12b is indeed carbon-rich as the
detection of HCN and C$_2$H$_2$ would suggest, or ii) the original
atmosphere was nearly carbon-rich and an efficient cloud formation
tipped it over in the carbon-rich regime (like in Fig. 9 in
\citealt{hell2014}, or iii) external irradiation (e.g cosmic rays)
drives a strong ion-neutral chemistry which leads to the emergence of
larger carbo-hydrate molecules (\citealt{rim2014}).

 \begin{acknowledgements}
{We highlight financial support of the European Community under the
  FP7 by an ERC starting grant number 257431. A summer scholarship for DT
  provided by the Royal Astronomical Society is highly acknowledged.}
 \end{acknowledgements}


\bibliography{bib}
\bibliographystyle{mn2e}

\newpage
\appendix
\section{Surface growth reactions and vapor pressure data}
The sets of surface reaction are provided here which have been applied
to model oxygen-rich and carbon-rich cloud particle growth by
gas-surface reactions as outlined in \cite{hell2008b}. The cloud
particle growth in a clearly oxygen-rich environment (e.g. C/O=0.43)
is described by the surface reaction listed in Table 1 in
\cite{hell2008b} plus those listed in Table~\ref{tab:Ogrowth} in the
present paper, adding to a total of 73 surface reactions. For the
carbon rich case, we use 84 surface reactions which include the
reactions for TiO$_2$[s], SiO$_2$[s], SiO[s], Fe[s], FeS[s], MgO[s],
and Al$_2$O$_3$[s] (32 reactions) as listed in Table 1 in
\cite{hell2008b} plus those given in Table~\ref{tab:Cgrowth}.

Table~\ref{tab:V0ci} contains the vapor pressure data used for the
materials that were added to our cloud model (C[s], MgS[s], TiC[s], SiC[s], KCl[s]). C[s], TiC[s], and SiC[s] are fits to the
JANAF table (1986). MgS[s] is taken from \cite{sharp1990}, and
KCl[s]  from Eq. (18) in \cite{mor2012}.

\begin{table*}
\caption{Monomer volumes $V_{\rm 0,s}$ of solid materials from
  and fit coefficients $c_i$ for the calculation of
  the saturation vapor pressures and difference Gibbs free
    energies. (*): done for this paper from JANAF tables (1986) -- (**): Sharp \& Huebner
  (1990), (***) Morley et al 2012.}
\label{tab:V0ci}
\centering
\begin{tabular}{r|c|r|r|r|r|r|lr}
condensate & $V_{\rm 0,s}$ [$10^{-23}$ cm$^{3}$] 
           & $c_1$ & $c_2$ & $c_3$ & $c_4$ & $c_5$ & fit of & \\
\hline
C[s]     & 1.002 & 1.01428E+6 & $-7.23043$E+5& 1.63039E02  & $-1.75890$E-3  &9.97416E08   & $\ln p_{\rm sat}$ & (*)\\
MgS[s]   & 3.296 & $-7.93442$E+4 & $-1.83584$E+5 & 6.33920 & $-8.89360$E-4
               & 0.0 &  $\Delta G^\prime$ & (**)\\
TiC[s]    & 2.018 & 1.11878E+5 & $-1.37612$E+6 & 3.20666E+2 &  $-4.63379$E-3 &  1.85306E-7 & $\Delta G^{\prime}$& (*)\\
SiC[s]    & 2.074 & 6.73337E+5 & $-1.24381$E+6 & 3.21779E+2 & $-4.54405$E-3 & 2.69711E-7 & $\Delta G^{\prime}$& (*)\\
KCl[s]    & 6.227 & \multicolumn{5}{l}{7.611 - 11382.0/T$_{\rm gas}$}   & $\log p_{\rm sat}$ & (***)
\end{tabular}
\end{table*}

\begin{table*}
\caption{Oxygen-rich cloud formation: Dust growth surface reactions for a oxygen-rich environment in addition to those listed in Table 1 in  \cite{hell2008b}. This set of 73 reactions is applied for all oxygen-rich models in the present paper.}
\label{tab:Ogrowth}
\centering
{\small
\begin{tabular}{ccll}
Index $r$ & Solid $s$     & Surface reaction                               & Key species \\\hline
61  & KCl[s]       &  KCl                            $\rightarrow$   KCl[s]  &   KCl\\
62  &  sylvite               &  KOH +  HCl       $\rightarrow$   KCl[s] +  H$_2$O &  min\{KOH, HCl\}\\
63  &                &  HK +  HCl                     $\rightarrow$   KCl[s] +  H$_2$  & min\{HK, HCl\}\\
64  &                &  KOH +  CaCl                   $\rightarrow$   KCl[s] +  CaOH & min\{KOH, CaCl\}\\
65  & MgS[s]           &  MgS                            $\rightarrow$   MgS[s] & MgS\\
66 &   niningerite              & 2 MgH + 2 FeS                    $\rightarrow$  2 MgS[s] +  H$_2$ + 2 Fe & min\{MgH, FeS\}\\
67 &                  & 2 MgH + 2 H$_2$S                 $\rightarrow$  2 MgS[s] + 3 H$_2$ & min\{MgH, H$_2$S\}\\
68 &                  & 2 MgOH + 2 H$_2$S                $\rightarrow$  2 MgS[s] + 2 H$_2$O + 1 H$_2$ &min\{MgOH, H$_2$S\}\\
69 &                  & MgOH +  CaS                   $\rightarrow$   MgS[s] +  CaOH & min\{MgOH, CaS\}\\
70 &                  &  MgO +  CS                     $\rightarrow$   MgS[s] +  CO & min\{MgO, CS\}\\
71 &                  & 2 MgH + 2 CS                     $\rightarrow$  2 MgS[s] +  H$_2$ + 2 C & min\{MgH, CS\} \\
72 &                  &  MgO +  H$_2$S                 $\rightarrow$   MgS[s] +  H$_2$O &min\{MgO, H$_2$S\}\\
73 &                  &  Mg +  H$_2$S                  $\rightarrow$   MgS[s] +  H$_2$ &min\{Mg, H$_2$S\} \\\hline
\end{tabular}
}
\end{table*}

\begin{table*}
\caption{Carbon-rich cloud formation: Dust growth surface reactions for a carbon-rich environment in addition to those listed in Table 1 in  \cite{hell2008b}
 for TiO$_2$[s], SiO$_2$[s], SiO[s], Fe[s], FeS[s], MgO[s], and Al$_2$O$_3$[s] (32 reactions). This set of 84 reactions is applied for all carbon-rich models.}
\centering
{\small
\begin{tabular}{ccll}
Index $r$ & Solid $s$     & Surface reaction              &    Key species            \\
\hline
33 & C[s]             &  C                             $\rightarrow$   C[s] & C\\
34 & carbon                &  C$_2$                         $\rightarrow$  2 C[s] & C$_2$ \\
35 &                 &  C$_3$                         $\rightarrow$  3 C[s]                     & C$_3$\\
36 &                 & 2 C$_2$H                        $\rightarrow$  4 C[s] +  H$_2$  & C$_2$H \\
37 &                 &  C$_2$H$_2$                $\rightarrow$  2 C[s] +  H$_2$     & C$_2$H$_2$\\
38 &                 &  CH$_4$                        $\rightarrow$   C[s] + 2 H$_2$     & CH$_4$\\
39 & TiC[s]           &  Ti +  C                      $\rightarrow$   TiC[s]                     & min\{Ti, C\} \\
40 &   titanium-               & 2 Ti +  C$_2$                  $\rightarrow$  2 TiC[s]  & min\{Ti, C$_2$\} \\
41 &   carbide             & 3 Ti +  C$_3$                  $\rightarrow$  3 TiC[s]      & min\{Ti, C$_3$\}   \\
42 &                 & 4 Ti + 2 C$_2$H                 $\rightarrow$  4 TiC[s] +  H$_2$  & min\{Ti, C$_2$H\} \\
43 &                 & 2 Ti +  C$_2$H$_2$          $\rightarrow$  2 TiC[s] +  H$_2$   & min\{Ti, C$_2$H$_2$\}  \\
44&                  &  Ti +  CH$_4$                 $\rightarrow$   TiC[s] + 2 H$_2$       &  min\{Ti, CH$_4$\} \\
45 &                 &  TiC                           $\rightarrow$   TiC[s]                           & TiC\\
46 &                 &  TiS +  C +  H$_2$           $\rightarrow$   TiC[s] +  H$_2$S   &  min\{TiS, C\} \\
47 &                 & 2 TiS +  C$_2$ + 2 H$_2$       $\rightarrow$  2 TiC[s] + 2 H$_2$S &  min\{TiS, C$_2$\}  \\
48 &                 & 3 TiS +  C$_3$ + 3 H$_2$       $\rightarrow$  3 TiC[s] + 3 H$_2$S  &  min\{TiS, C$_3$\}\\
49 &                 & 4 TiS + 2 C$_2$H + 3 H$_2$      $\rightarrow$  4 TiC[s] + 4 H$_2$S &  min\{TiS, C$_2$H\} \\
50 &                 & 2 TiS +  C$_2$H$_2$ +  H$_2$  $\rightarrow$  2 TiC[s] + 2 H$_2$S  &  min\{TiS, C$_2$H$_2$\} \\
51 &                & TiS +  CH$_4$ +  H$_2$      $\rightarrow$   TiC[s] +  H$_2$S &  min\{TiS, CH$_4$\} \\
52 &                 &  TiC$_2$                       $\rightarrow$   TiC[s] + C & TiC$_2$\\
53 & SiC[s]           &  Si +  C                    $\rightarrow$   SiC[s] &  min\{Si, C\} \\
54 &  silicon-                & 2 Si +  C$_2$                $\rightarrow$  2 SiC[s] & min\{Si, C$_2$\}  \\
55 &  carbide                & 3 Si +  C$_3$                  $\rightarrow$ 3 SiC[s] & min\{Si, C$_3$\} \\
56 &                  & 4 Si + 2 C$_2$H                 $\rightarrow$  4 SiC[s] +  H$_2$ & min\{Si, C$_2$H\}\\
57 &                  & 2 Si +  C$_2$H$_2$             $\rightarrow$  2 SiC[s] +  H$_2$ & min\{Si, C$_2$H$_2$\} \\
58 &                 & Si +  CH$_4$                 $\rightarrow$   SiC[s] + 2 H$_2$ & min\{Si, CH$_4$\} \\
59 &                 &  SiC                           $\rightarrow$   SiC[s] & SiC\\
60 &                  &  SiS +  C + H$_2$           $\rightarrow$   SiC[s] +  H$_2$S & min\{SiS, C\}  \\
61 &                  & 2 SiS +  C$_2$ + 2 H$_2$       $\rightarrow$  2 SiC[s] + 2 H$_2$S& min\{SiS, C$_2$\}  \\
62  &                  & 3 SiS +  C$_3$ + 3 H$_2$       $\rightarrow$  3 SiC[s] + 3 H$_2$S & min\{SiS, C$_3$\} \\
63 &                  & 4 SiS + 2 C$_2$H + 3 H$_2$      $\rightarrow$  4 SiC[s] + 4 H$_2$S & min\{SiS, C$_2$H\}  \\
64 &                  & 2 SiS +  C$_2$H$_2$ + H$_2$  $\rightarrow$  2 SiC[s] + 2 H$_2$S & min\{SiS, C$_2$H$_2$\}  \\
65 &                  &  SiS +  CH$_4$ +  H$_2$      $\rightarrow$   SiC[s] +  H$_2$S & min\{SiS, CH$_4$\} \\
66 &                 &  Si$_2$C                       $\rightarrow$   SiC[s] +  Si & Si$_2$C\\
67 &                 &  SiC$_2$                       $\rightarrow$   SiC[s] + C & SiC$_2$ \\
68 &                  &  SiH$_4$ +  C              $\rightarrow$   SiC[s] + 2 H$_2$ & min\{SiH$_4$, C\}   \\
69 &                 & 2 SiH$_4$ +  C$_2$             $\rightarrow$  2 SiC[s] + 4 H$_2$ & min\{SiH$_4$, C$_2$\}\\
70 &                  & 3 SiH$_4$ +  C$_3$             $\rightarrow$  3 SiC[s] + 6 H$_2$  & min\{SiH$_4$, C$_3$\}\\
71 &                  & 3 SiH$_4$ + 2 C$_2$H           $\rightarrow$  4 SiC[s] + 9 H$_2$ & min\{SiH$_4$, C$_2$H\} \\
72 &                  & 2 SiH$_4$ +  C$_2$H$_2$        $\rightarrow$  2 SiC[s] + 5 H$_2$ & min\{SiH$_4$, C$_2$H$_2$\} \\
73 &                  &  SiH$_4$ +  CH$_4$            $\rightarrow$  SiC[s] + 4 H$_2$ & min\{SiH$_4$, CH$_4$\}  \\
74 &  KCl[s]           &  KCl                           $\rightarrow$   KCl[s]  & KCl\\
75 &   sylvite               &  HK +  HCl                    $\rightarrow$  KCl[s] +  H$_2$ & min\{KH, HCl\} \\
76 &                  & 2 CKN + 2 H$_2$O +  CaCl$_2$   $\rightarrow$  2 KCl[s] + 2 C +  N$_2$ +  Ca(OH)$_2$ +  H$_2$  & min\{CKN, CaCl$_2$\}\\
77 & MgS[s]           &  MgS                           $\rightarrow$   MgS[s] & MgS\\
78 &  niningerite                 & 2 MgH + 2 H$_2$S                $\rightarrow$  2 MgS[s] + 3 H$_2$ & min\{MgH, H$_2$S\} \\
79 &                  & 2 MgOH + 2 H$_2$S               $\rightarrow$  2 MgS[s] + 2 H$_2$O +  H$_2$& min\{MgOH, H$_2$S\}   \\
80 &                  &  MgOH +  CaS                  $\rightarrow$   MgS[s] +  CaOH & min\{MgOH, CaS\}  \\
81 &                  & 2 MgN + 2 H$_2$S                $\rightarrow$  2 MgS[s] + 2 H$_2$ +  N$_2$ & min\{MgN, H$_2$S\}  \\
82 &                  & 2 MgH + 2 CS                    $\rightarrow$  2 MgS[s] +  H$_2$ + 2 C & min\{MgH, CS\}  \\
83 &                  & 2 MgOH + 2 TiS                  $\rightarrow$  2 MgS[s] +  H$_2$ + 2 TiO & min\{MgOH, TiS\}  \\
84 &                  &  Mg +  H$_2$S                 $\rightarrow$   MgS[s] +  H$_2$ & min\{Mg, H$_2$S\}  \\
                 \hline
\label{tab:Cgrowth}
\end{tabular}
}
\end{table*}

\section{More detailed results}
\begin{figure*}
 \includegraphics[scale=0.3]{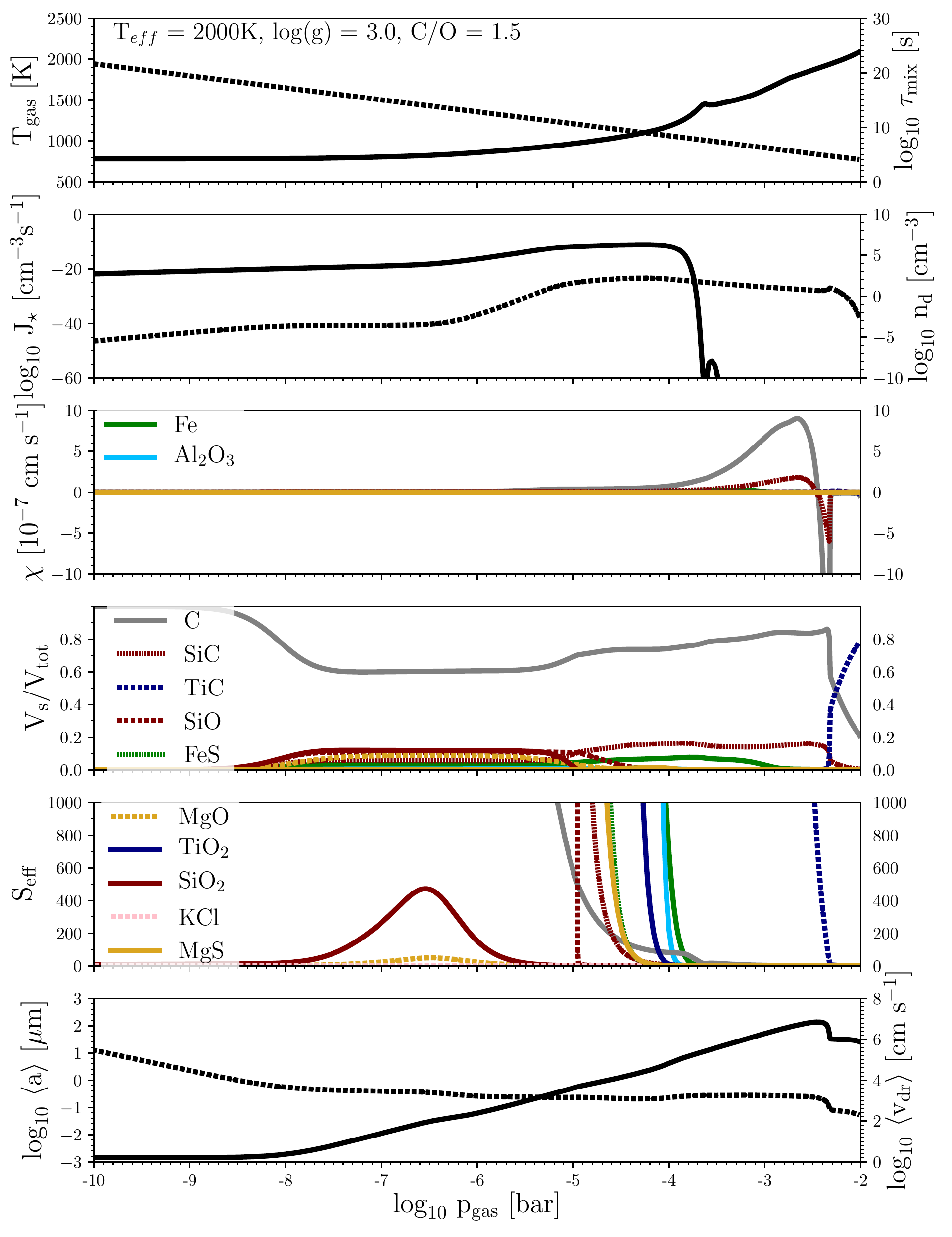}
\includegraphics[scale=0.3]{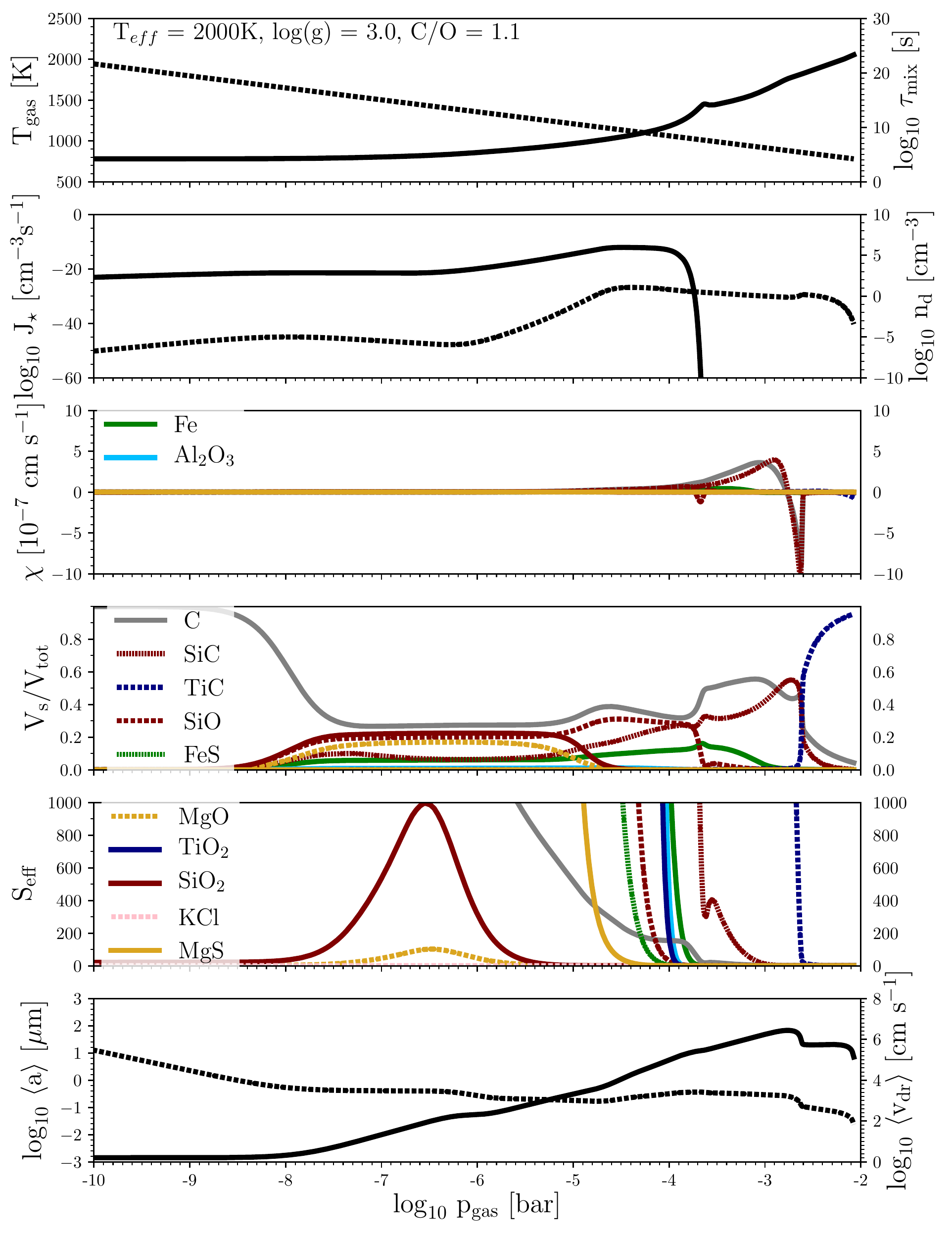}
\includegraphics[scale=0.3]{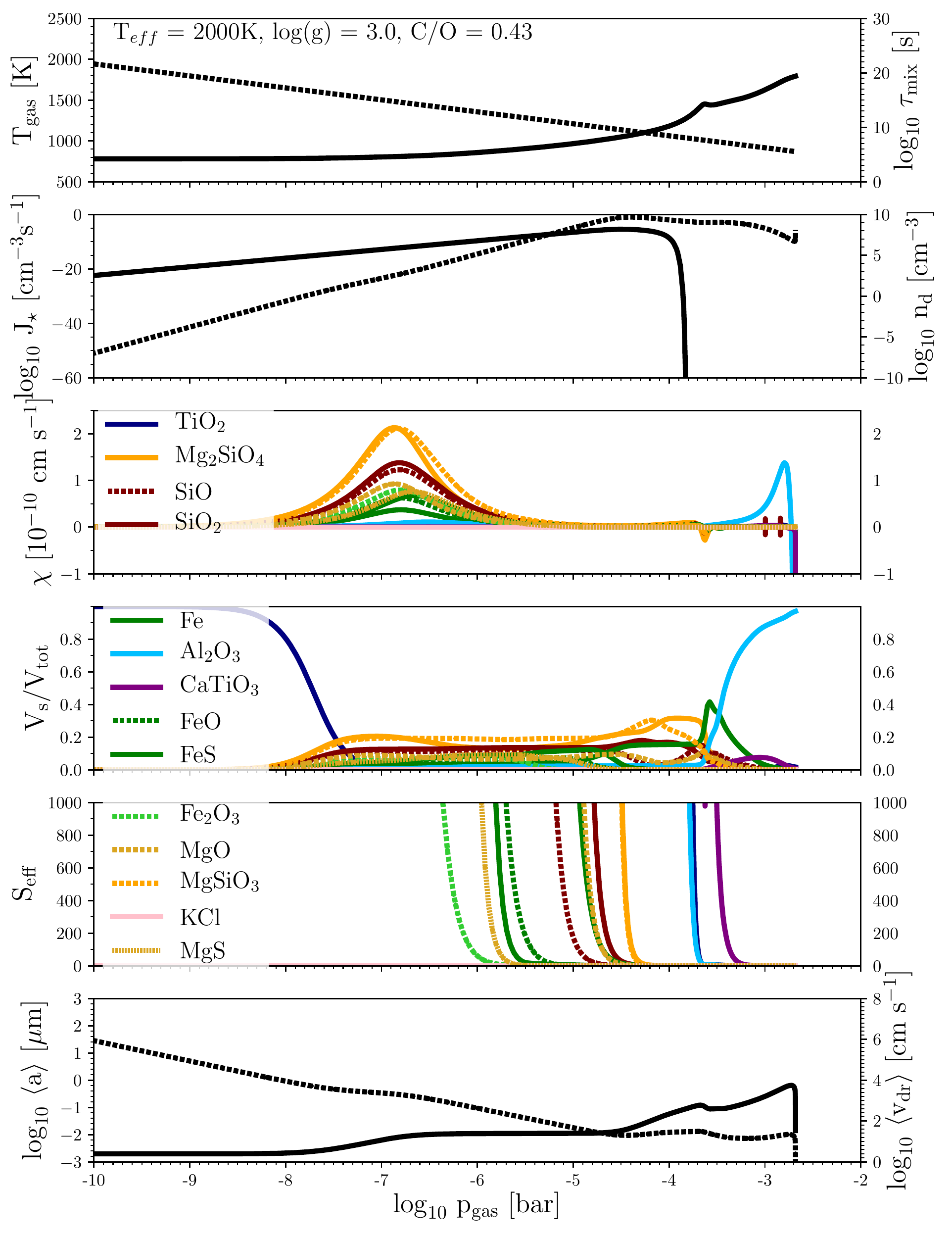}
\caption{Detailed information about the cloud structure forming in an atmospheric gas with initial C/O ratios of 0.43, 1.1, 1.5 and a pre-scribed {\sc Drift-Phoenix} atmosphere profile for T$_{\rm eff}=2000$K and log(g)=3.0. {\bf 1st rows:} local gas temperature T$_{\rm gas}$ [K] (solid, left), mixing time scale $\tau_{\rm mix}$ [s] (dashed, right); {\bf 2nd rows:} seed formation rate $J_*$ [cm$^{-3}$s$^{-1}] (solid, left)$, number density of cloud particles n$_{\rm d}$ [cm$^{-3}$] (right, dashed),  {\bf 3rd rows:} net growth velocity for individual materials $s$; {\bf 4th rows:} volume fraction $V_{\rm s}/V_{\rm tot}$ for material $s$;  {\bf 5th rows:} effective supersaturation ratio S${\rm eff}$ for each material $s$;  {\bf 6th rows:} mean cloud particles radius $\langle a \rangle$ [$\mu$m], drift velocity v$_{\rm drift}$ [cm s$^{-1}$] with respect to the local $\langle a \rangle$. The colour code is the same for all panels. All quantities are plotted versus the local gas pressure p$_{\rm gas}$ [bar].
}
 \label{fig:CtoOchange_individual}
\end{figure*}

Figure~\ref{fig:CtoOchange_individual} provides more detailed results of the cloud structures for the initial C/O ratios 0.43, 1.1, 1.5 the results of which were presented in Sect.~\ref{CtoOdiff}. The plots in Fig.~\ref{fig:CtoOchange_individual} have the same set-up like in our previous publication to allow for a sensible comparison of cloud structures forming in different atmospheric conditions. The plots allow for a more detailed comparison between the oxygen-rich and the carbon-rich cloud formation.

\end{document}